\documentclass[twocolumn,twocolappendix]{aastex63}
\usepackage{amsmath}
\usepackage{soul}

\newcommand\kms{km s$^{-1}$}

\newcommand\teff{$T_{\rm eff}$}
\newcommand\logg{$\log g$}

\defcitealias{kounkel2023}{Paper I}
\usepackage{listings}

\begin{document}
\title{ABYSS II: Identification of young stars in optical SDSS spectra and their properties}

\author[0009-0004-9592-2311]{Serat Saad}
\affil{Department of Physics and Astronomy, Vanderbilt University, VU Station 1807, Nashville, TN 37235, USA}
\author{Kaitlyn Lane}
\affil{Department of Physics and Astronomy, Vanderbilt University, VU Station 1807, Nashville, TN 37235, USA}
\author[0000-0002-5365-1267]{Marina Kounkel}
\affil{Department of Physics and Astronomy, University of North Florida, 1 UNF Dr., Jacksonville, FL 32224, USA}
\affil{Department of Physics and Astronomy, Vanderbilt University, VU Station 1807, Nashville, TN 37235, USA}
\email{marina.kounkel@unf.edu}
\author[0000-0002-3481-9052]{Keivan G.\ Stassun}
\affil{Department of Physics and Astronomy, Vanderbilt University, VU Station 1807, Nashville, TN 37235, USA}
\author[0000-0002-7795-0018]{Ricardo L\'opez-Valdivia}
\affil{Universidad Nacional Aut\'onoma de M\'exico, Instituto de Astronom\'ia, AP 106,  Ensenada 22800, BC, M\'exico}
\author[0000-0001-6072-9344]{Jinyoung Serena Kim}
\affiliation{Steward Observatory, Department of Astronomy, University of Arizona, 933 North Cherry Avenue, Tucson, AZ 85721, USA}

\author[0000-0002-5855-401X]{Karla Pe\~na Ram\'irez}
\affiliation{Centro de Astronomía (CITEVA), Universidad de Antofagasta, Av. Angamos 601, Antofagasta, Chile}

\author[0000-0003-1479-3059]{Guy S. Stringfellow}
\affiliation{Center for Astrophysics and Space Astronomy, Department of Astrophysical and Planetary Sciences, University of Colorado 389-UCB, Boulder, CO 80309, USA}

\author[0000-0001-8600-4798]{Carlos G. Rom\'an-Z\'u\~niga}
\affiliation{Universidad Nacional Aut\'onoma de M\'exico, Instituto de Astronom\'ia, AP 106,  Ensenada 22800, BC, M\'exico}

\author[0000-0001-9797-5661]{Jes\'us Hern\'andez}
\affiliation{Universidad Nacional Aut\'onoma de M\'exico, Instituto de Astronom\'ia, AP 106,  Ensenada 22800, BC, M\'exico}

\author[0000-0002-0826-9261]{Scott J. Wolk}
\affiliation{Center for Astrophysics, Harvard \& Smithsonian, 60 Garden St., Cambridge MA. 02138, USA}

\author{Lynne A. Hillenbrand}
\affiliation{Department of Astronomy, California Institute of Technology, Pasadena, CA 91125, USA}

\begin{abstract}
We developed a tool that measures equivalent widths of various lines in low resolution optical spectra, and it was applied to stellar spectra obtained as part of SDSS-V and LAMOST programs. 
These lines, such as Li I which directly indicates stellar youth, or optical H I and Ca II which in emission indicate activity associated with stellar youth, are commonly seen in YSOs.
We observe several notable differences in the properties of these lines between YSOs and the field stars. Using these data, we devise a set of criteria through which it is possible to confirm the youth of stars that have been observed by the ABYSS program, as well as to identify likely young stars that have serendipitously been observed by other programs. We examine the decrement of H lines seen in emission in CTTSs, and estimate the properties of the accretion stream that is responsible for the production of these lines. Finally, we examine the evolution of Li I as a function of age, and characterize the scatter in its abundance that appears to be intrinsic in young M dwarfs.
\end{abstract}


\section{Introduction}

In its observing strategy, Sloan Digital Sky Survey in its fifth iteration (SDSS-V) is set to obtain optical and near IR spectra of several million stars through the Milky Way Mapper (MWM) program \citep{almeida2023}. This will be achieved with both the BOSS (Baryonic Oscillation Spectroscopic Survey) and APOGEE (APO Galactic Evolution Experiment) spectrographs, using a state of the art fiber robotic positioner \citep{pogge2020} that allows the simultaneous observation of up to 500 targets. As part of the SDSS-V MWM program the APOGEE \& BOSS Young Star Survey (ABYSS) is expected to produce multi-epoch spectra of $>$100,000 photometrically identified young star candidates with ages $<$30 Myr \citep[hereafter Paper I]{kounkel2023}.

To-date, single epoch spectra of several tens of thousands of such candidates have been observed, allowing us to perform an initial characterization of their properties. Since the initial selection of the ABYSS targets was not free of contamination, it is necessary to use spectroscopic criteria to separate out older field stars from the bona fide young stellar objects (YSOs) to enable any follow up work. We take advantage of the fact that optical spectra of young stars typically exhibit a number of unique features (e.g., Li I absorption, H$\alpha$ emission, inflated radii) that are not present in the field stars, allowing us to perform their identification.

The Li I 6708 \AA\ absorption line has particular importance. Li I is present in the ISM, and as such, it is included in the chemical composition of YSOs as they form. However Li is easily destroyed: as soon as the internal temperature of a star reaches $3\times10^6$ K, it is rapidly processed in nuclear reactions \citep{clayton1983}. In higher mass stars with radiative envelopes, a trace fraction of Li I will persist near the photosphere regardless of the internal temperature. However, when fully or mostly convective stars reach the age where they have sufficiently contracted to reach that temperature in the core/convective boundary layer, Li I becomes depleted everywhere within its envelope. Thus, Li I will not be detected in low mass field stars, and its detection can act as an unambiguous confirmation of stellar youth \citep[e.g.,][]{briceno1997,jeffries2014,gutierrez-albarran2020}
although some care needs to be applied with regards to the solar-type stars \citep{zerjal2019}. Additionally, at its strongest, Li I absorption typically has equivalent widths of only 0.5 \AA, and this feature is found near several Fe I lines. Thus, care needs to be taken in order to minimize the confusion and contamination.

In the case of YSOs with dusty disks, they often exhibit strong and wide emission lines from the accretion stream shock and outflows. The strongest of those lines in the optical spectra is H$\alpha$, often presenting equivalent widths (EqW) of tens to hundreds of \AA\ \citep{white2003}. Particularly energetic shocks are capable of exciting other transitions of H \citep{kwan2011,wilson2022,campbell2023}, as well as a number of other elements, including various O, Fe, He, S, N, and others \citep{hamann1992,hamann1992a,hamann1992b,hamann1994,baldovin-saavedra2012,ballabio2022}. Young stars exhibiting accretion are known as Classical T Tauri Stars (CTTSs). Since disks tend to be short-lived, CTTSs are not ubiquitous among all young stars. Broad emission lines can help to cleanly separate out YSOs from the field stars, although a few rare types of more evolved stars (e.g, cataclysmic variables) could also exhibit them \citep{echevarria1988}. Additionally, in some rare cases H$\alpha$ in CTTSs can be so strong, it may confuse the data processing pipelines in large surveys such as SDSS, which could erroneously result in masking such a feature as a cosmic ray.

Young stars also tend to be $10^2-10^4$ times more magnetically active than the field stars \citep[e.g.,][]{skumanich1972,feigelson2007,kounkel2022a}. This activity produces a number of emission lines, most notably in H$\alpha$ and other H lines, as well as Ca \citep{vaughan1978,white2003,briceno2019}, although they are significantly weaker and narrower than the lines dominated by accretion. Young stars with emission lines driven primarily from activity are known as Weak Lined T Tauri Stars (WTTSs). Activity-driven emission is significantly longer lived however, it is able to persist for several Gyr in late M dwarfs, and and for several 100s of Myr in early M dwarfs \citep{west2008,newton2017}. However, any emission in G \& K dwarfs is a clearer indication of stellar youth.

In addition to the above features that are primarily observed in the optical regime, with sufficiently high signal to noise, \logg\ can also be used as a reliable tracer in many young stars. Low mass pre-main sequence (PMS) stars are still inflated over their main sequence counterparts, as such they have a systematically lower \logg\ in comparison to the field stars. For high mass stars, it is the inverse: because OB (and to a lesser extent A) stars evolve so rapidly, catching them on the main sequence is more likely than not to signify their youth in comparison to those stars that have already evolved into giants. Although there has been difficulty in accurately calibrating \logg\ measurements of young stars in the past, recent efforts to process SDSS spectra (both APOGEE \& BOSS) now allow \logg\ of low mass YSOs to be an independent estimate of age \citep{olney2020,sprague2022}. 

Each individual tracer may provide a strong indication of stellar youth, and, for some sources, particularly if the scope of the study is limited to e.g., an individual star forming region, simple criteria are capable of producing clean sample. However, any individual criteria are significantly easier to apply to select low mass KM dwarfs, in comparison to OBAFG stars, and stars with the age of a few Myr are easier to identify than those with an age of 20-30 Myr.

Dealing with the data from a large all-sky spectroscopic survey results in its own challenges. Given that YSOs tend to be somewhat rare, any selection criteria have to be carefully adjusted, as even a small fraction of the field stars in the the parameter space similar to that of the young stars can overwhelm the sample, while any cuts that are too strict may exclude a significant fraction of bona fide YSOs that could have been easily identifiable in a smaller study. As such, using a combination of different criteria is advantageous for developing a cleaner but still comprehensive sample. At the same time, however, it is necessary to efficiently and carefully extract relevant spectral features in a manner that could be applied to all of the stars regardless of their mass, evolutionary status, or the unique properties than a star may exhibit.

In this paper we aim to develop a pipeline for measuring equivalent widths of various youth-sensitive lines in low resolution optical spectra, such as those produced with BOSS and LAMOST. Using these data, we develop a classifier for identification of young stars, and we study the evolution of the properties of these lines as a function of age.

\section{Data} \label{sec:data}


BOSS is a low resolution optical spectrograph with $R\sim$1800 with the typical pixel scale of $\sim$1 \AA, covering a wavelength range of 3600--10400 \AA \citep{smee2013}. Twin instruments are mounted at two observatories, at Apache Point Observatory \citep[APO;][]{gunn2006,blanton2017} and Las Campanas Observatory \citep[LCO;][]{bowen1973}, to ensure a complete coverage of the entire sky. Each spectrograph is capable of observing up to 500 spectra within 3$^\circ$ and 2$^\circ$ field of view respectively. In prior iterations of SDSS, BOSS was primarily to observe extragalactic targets, obtaining spectra of only a few thousand of stars in total \citep{abdurrouf2022}, although it did include a couple of fields rich in PMS stars \citep{suarez2017}. Since the transition to SDSS-V in 2021, it has observed well over 400,000 stars, including more than 17,000 stars targeted by ABYSS.

LAMOST (Large Sky Area Multi-Object Fiber Spectroscopic Telescope) is a spectrograph very similar to BOSS in terms of its resolution, although it does have a slightly narrower wavelength range of 3700--9000 \AA, and it can observe up to 4000 stars simultaneously \citep{yan2022}. Beginning its operations in 2011, it has obtained spectra of more than 10 million stars. So far there have been only a few studies utilizing LAMOST specta of young stars \citep{{liu2021,wang2022c,hernandez2023,lin2023}}, however, in LAMOST DR8 there have been more than 13,000 serendipitously observed stars that are included in ABYSS SDSS targeting (most of which to-date are yet to be observed with BOSS). Given the similarity between BOSS \& LAMOST, it is beneficial include these data in the analysis.

Fundamental stellar parameters for the data from both of these spectrographs have been extracted using BOSS Net (Sizemore L. et al. submitted), a neural net that is capable of predicting self-consistent \teff\ and \logg\ (with respective precision of 0.008 dex and 0.1 dex at SNR$\sim$15) values that are calibrated to theoretical models for stars of all types, including PMS stars, main sequence sources, brown dwarfs, red giants, subdwarfs, white dwarfs, as well as OB stars. Additionally, it provides an independent estimate of stellar radial velocities (RV). This is important, since RVs that are included in LAMOST data releases have a number of artefacts, systematic offsets on order of up to 10 \kms\ that vary depending on sky position and spectral type. Accurate RVs are crucial for accurate centroiding of spectral lines, especially those that are weak and narrow.

\section{Methods}\label{sec:analysis}

\subsection{Measuring line widths}

\begin{figure}
\epsscale{1.1}
\plotone{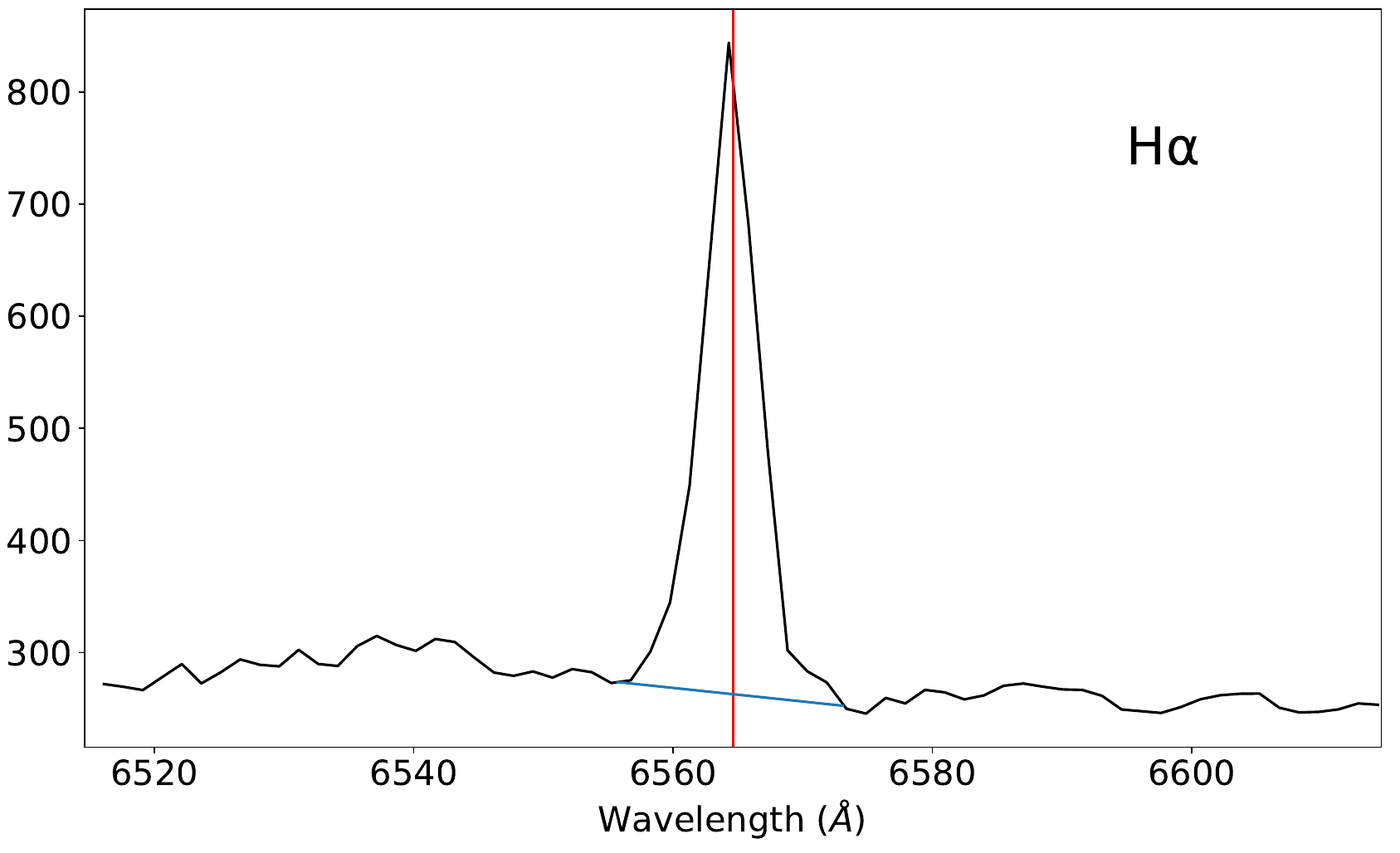}
\plotone{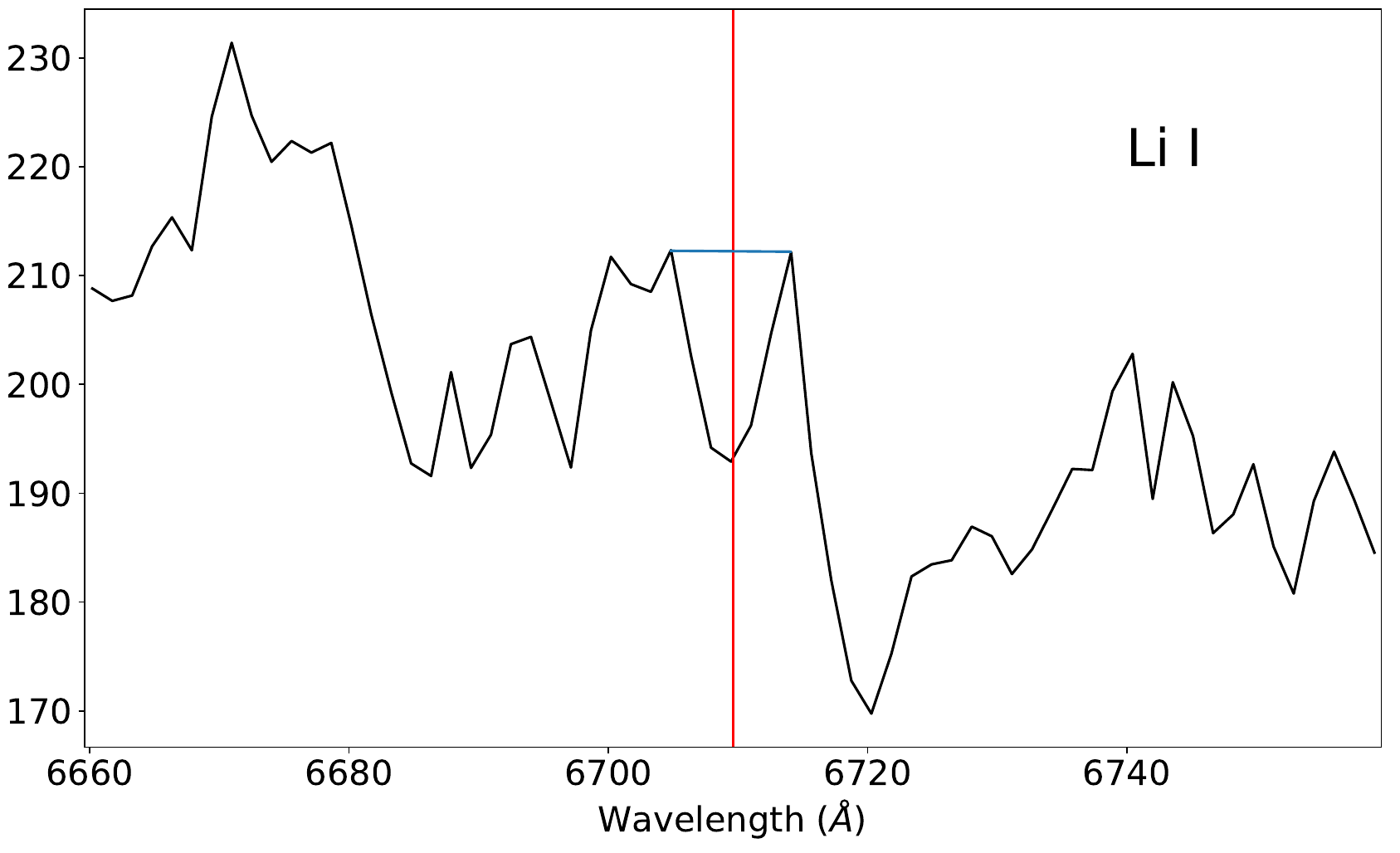}
\caption{An example of a spectral lines (H$\alpha$ and Li I) of a PMS star in BOSS spectra. The red line is centered on the feature, the blue line is the manually defined continuum over which the line is integrated to measure EqW.
\label{fig:manual}}
\end{figure}

Here we describe the development of the LineForest pipeline which extracts a number of youth sensitive features from the optical spectra.

\subsubsection{Initial measurements}

Different lines present different challenges in the ability of measuring accurate EqW. Some lines are quite strong, such as H$\alpha$, with absolute widths (AbW) that can vary with spectral type, and depending on the range over which the line is integrated, the EqW would either be underestimated, or it could be contaminated by the flux well outside of the line. On the other hand, Li I is a relatively weak absorption line; typically not exceeding EqW of 0.5 \AA \citep{briceno1997}. A number of lines such as Fe I and CN can often blend in together with Li I when its strength decreases to $<$0.2 \AA. One of the methods through which Li I EqW has been measured in a large census in the past was through identifying the closest matching spectrum of an evolved star and subtracting it out such that only Li I remains \citep{zerjal2019}, however while it is possible to do this in high resolution and high SNR spectra, such an approach creates a challenge for BOSS, particuarly because $>70$\% of visits have SNR$<$30. We attempted a number of other approaches - Gaussian profile fitting, automated continuum determination, automated width estimation -- all of which had some degree of success in a subset of stars, but none could be generalized across all sources even within a single class.

We found that the most reliable approach to measuring EqW was through manually defining each line in each spectrum, which is an adaptive approach that allows to measure the width of the entire line, and enables skipping sources that do not present a feature in either emission or absorption. This task is not feasible to do for all stars observed in a large survey due to a significant person-power requirement. However, doing it on a subset of stars can create a vetted set of training labels for data driven machine learning applications which can then be packaged into an automated pipeline.

For this purpose, we selected a random set of $\sim$3500 spectra, among which about 1000 sources targeted by ABYSS, and the remaining were chosen to create a representative subset of targets from other programs. Those other programs include OB stars, brown dwarfs, white dwarfs, cataclysmic variables, and red giants, featuring $\sim$250 stars in each class. For all those sources we have measured line properties of the 23 most prominent transitions, while another 29 lines were measured in a subset of 350 spectra to enable the model to learn the shape of the continuum at these transitions. This represents a total of 52 lines (out of which a half are Balmer or Paschen H series lines that fall in the BOSS wavelength range). The features are listed in Table \ref{tab:lines}.

Using a custom-built interactive code previously used in \citet{campbell2023}, we recorded, for each line in each source, the wavelength and flux ($\lambda_1$, $F_1$) near the start of the line as well as those near the end of the line ($\lambda_2$, $F_2$). These set of points define the continuum under which the spectrum was integrated to estimate EqW (Figure \ref{fig:manual}). By convention, absorption lines are defined with positive EqW, and emission lines are defined with negative EqW. The difference $\lambda_2-\lambda_1$ corresponds to AbW. In cases where no apparent line was observed, both EqW and AbW were recorded as 0. In cases where significant blending was suspected, depending on the exact profile the line was either recorded as 0, or the continuum was drawn over the portion of the line most likely to be associated with the given element. While such approach is imperfect, the neural net is capable of interpolating through these measurements to achieve a more general solution.

We do note that the measured line properties do not take into account various physical effects that may alter them, for example veiling due to accretion, and model fitting is required in order to characterize it \citep[e.g.,][]{fang2020a,fang2021}. We defer full analysis of the effects of veiling in young stars to future works in the series.

Low resolution spectra such as BOSS may present a challenge in identifying some weaker lines if the radial velocity is very uncertain, particularly for sources with very low SNR. As a consequence of this, if during manual measurements a confident identification was not possible, nothing would be recorded for it, thus attempting to condition the neural net to disregard it.

However, it should be noted that in some cases, some lines may be systematically misattributed. For example, He transitions are most commonly observed in OB-type stars, although they can also be seen in emission in low mass PMS stars as a result of certain processes related to accretion. For instance, in a number of low mass main sequence stars, the Fe I 6678 \AA\ absorption feature has almost identical wavelength to He I 6678 line. It is unlikely that the lines are blended in single stars as they are prominent at completely different \teff\ regimes, however, for the sake of self-consistency, all of the likely misattributed measurements were preserved in the sample.

In some cases there is considerable blending, for example, Pa 13, Pa 15, and Pa 16 are located extremely close to the lines of Ca II triplet. Significant confusion is also  presetnt between Ca H and H$\epsilon$ lines. While some effort was made to define the continuum in such a way as to minimize the contribution from the neighboring strong features. Particularly, in young low mass stars, Calcium lines are in emission, with only weak Pa lines. It was not always possible to separate the lines cleanly, and thus in some of the approaches described below, there are cases with partial confusion among those lines. Reasonable judgement should be exercised in using the catalog for the appropriate \teff\ ranges for each line, and referring back to the spectra of the sources is suggested for the unexpected features.


\begin{deluxetable*}{cccccccl}
\tablecaption{Set of lines used in this study
\label{tab:lines}}
\tabletypesize{\scriptsize}
\tablewidth{\linewidth}
\tablehead{
  \colhead{Line} &
  \colhead{$\lambda$ \tablenotemark{$a$}} &
  \colhead{$\Delta\lambda$} &
  \colhead{Precision} &
  \colhead{Recall} &
  \colhead{EqW $\sigma$} &
  \colhead{AbW $\sigma$}  & 
  \colhead{Notes \tablenotemark{$c$}}\\
  \colhead{} &
  \colhead{(\AA)} &
  \colhead{(\AA)} &
  \colhead{} &
  \colhead{} &
  \colhead{(dex)\tablenotemark{$^b$}} &
  \colhead{(dex)\tablenotemark{$^b$}} &
  \colhead{}
  }
\startdata
\hline
H$\alpha$		&	6562.8		&  200 	& 0.98 & 0.96	  &  0.08  &  0.09 & \\
H$\beta$ 		&	4861.3		&  200 	& 0.98 & 0.94	  &  0.09  &  0.1 & \\
H$\gamma$ 		&	4340.5		&  200 	& 0.95 & 0.93	  &  0.1  &  0.12 & \\
H$\delta$ 		&	4101.7		&  200 	& 0.94 & 0.94	  &  0.1  &  0.11 & \\
H$\epsilon$ 	&	3970.1		&  200  & 0.95 & 0.94	  &  0.08  &  0.1 & [E]; Ca II H blend \\
H8 				&	3889.064	&  200 	& 0.91 & 0.66	  &  0.08  &  0.06 & \\
H9 				&	3835.391	&  200 	& 0.91 & 0.71	  &  0.11  &  0.1 & \\
H10 			&	3797.904	&  200 	& 0.82 & 0.77	  &  0.12  &  0.08 & \\
H11 			&	3770.637	&  200 	& 0.9 & 0.72	  &  0.1  &  0.08 & \\
H12 			&	3750.158	&  50 	& 0.85 & 0.79	  &  0.14  &  0.09 & \\
H13 			&	3734.369	&  50 	& 0.77 & 0.65	  &  0.13  &  0.09 & \\
H14 			&	3721.945	&  50 	& 0.87 & 0.68	  &  0.09  &  0.07 & \\
H15 			&	3711.977	&  50 	& 0.87 & 0.81	  &  0.12  &  0.15 & \\
H16 			&	3703.859	&  50 	& 0.88 & 0.7	  & &              & \\
H17 			&	3697.157	&  50 	& 0.56 & 0.38	  &  0.11  &  0.2  & \\
Pa7 			&	10049.4889	&  200 	& 0.9 & 0.84	  &  0.13  &  0.14 & \\
Pa8 			&	9546.0808	&  200 	& 0.93 & 0.9	  &  0.12  &  0.11 & \\
Pa9 			&	9229.1200	&  200 	& 0.92 & 0.85	  &  0.11  &  0.12 & \\
Pa10 			&	9014.909	&  200 	& 0.83 & 0.72	  &  0.16  &  0.17 & \\
Pa11 			&	8862.782	&  200 	& 0.86 & 0.78	  &  0.08  &  0.12 & \\
Pa12 			&	8750.472	&  200 	& 0.86 & 0.92	  &  0.11  &  0.11 & \\
Pa13 			&	8665.019	&  200 	& 0.73 & 1.0	  &  0.07  &  0.08 & [E]; Ca II 8662.140 \AA\ blend \\
Pa14 			&	8598.392	&  200 	& 0.75 & 0.9	  &  0.11  &  0.08 & \\
Pa15 			&	8545.383	&  200 	& 0.83 & 0.83	  & &              & [E]; Ca II 8542.089 \AA\ blend \\
Pa16 			&	8502.483	&  200 	& 0.62 & 1.0	  & &              & [E]; Ca II 8498.018 \AA\ blend \\
Pa17 			&	8467.254	&  200 	& 0.73 & 0.8	  & &              & \\
Ca II 			&	8662.140	&  50   & 0.99 & 0.96	  &  0.06  &  0.07 & [L]; Pa 13 blend \\
Ca II 			&	8542.089	&  50   & 0.98 & 0.98	  &  0.06  &  0.07 & [L]; Pa 15 blend \\
Ca II 			&	8498.018	&  50   & 0.97 & 0.94	  &  0.07  &  0.07 & [L]; Pa 16 blend \\
Ca II K 		&	3933.6614	&  200  & 0.94 & 0.92	  &  0.08  &  0.08 & \\
Ca II H 		&	3968.4673	&  200  & 0.96 & 0.92	  &  0.07  &  0.08 & [L]; H$\epsilon$ blend \\
He I 			&	6678.151	&  50   & 0.92 & 0.83	  &  0.13  &  0.08 & [E]; Fe I 6678 \AA\ blend \\
He I 			&	5875.621	&  50   & 0.71 & 0.55	  &  0.13  &  0.1  & [E] \\
He I 			&	5015.678	&  50   & 0.43 & 0.51	  &  0.15  &  0.17 & [E]; Fe II 5018 \AA\ blend \\
He I 			&	4471.479	&  50   & 0.81 & 0.62	  &  0.14  &  0.11 & [E]  \\
He II 			&	4685.7		&  50   & 0.08 & 0.38	  & &              & [E] \\
N II 			&	6583.450	&  50   & 0.74 & 0.66	  &  0.19  &  0.11 & \\
N II 			&	6548.050	&  50   & 0.65 & 0.53	  &  0.21  &  0.12 & \\
S II 			&	6716.440	&  50   & 0.81 & 0.74	  &  0.12  &  0.08 & \\
S II 			&	6730.816	&  50   & 0.69 & 0.59	  &  0.16  &  0.1  & \\
Fe II 			&	5018.434	&  50   & 0.76 & 0.85	  &  0.15  &  0.16 & \\
Fe II 			&	5169.030	&  50   & 0.81 & 0.81	  &  0.09  &  0.08 & \\
Fe II 			&	5197.577	&  50   & 0.57 & 1.0	  & &              & \\
Fe II 			&	6432.680	&  50   & 0.71 & 0.5	  & &              & \\
O I 			&	5577.339	&  50   & 0.8 & 0.53	  &  0.12  &  0.06 & \\
O I 			&	6300.304	&  50   & 0.47 & 0.56	  &  0.19  &  0.11 & \\
O I 			&	6363.777	&  50   & 0.2 & 0.25	  & &              & \\
O II 			&	3727.42		&  50   & 0.82 & 0.53	  &  0.23  &  0.17 & [E]; Fe I 3727 \AA\ blend \\
O III 			&	4958.911	&  50   & 0.5 & 0.33	  & &              & [E] \\
O III 			&	5006.843	&  50   & 0.78 & 0.78	  &  0.16  &  0.06 & [E] \\
O III 			&	4363.85		&  50   & 1.0 & 0.33	  & &              & [E] \\
Li I 			&	6707.760	&  50   & 0.88 & 0.64	  &  0.12  &  0.09 & [L]; Fe I 6707 \AA\ blend \\
\hline
\enddata
\tablenotetext{a}{All wavelengths are given in air, typically taken from \citet{nist}.}
\tablenotetext{b}{Scatter between the manual measurements and the predictions; reported only if the test set has $>$10 spectra with EqW$>$0.2 \AA}
\tablenotetext{c}{E: Line is most prominent in early type stars, L: Line is most prominent in late type stars}
\end{deluxetable*}

\subsubsection{Model architecture \& Training}

We categorize the studied lines into two groups: one contains lines that are typically broad, and the other groups those lines that are typically narrow. Broad lines (which include most of H lines, as well as Ca H \& K) are evaluated over a $\pm$200 \AA\ window centered on the doppler-corrected line. Narrow lines (all of the remaining elements, as well as high order lines in the Balmer series, particularly those found close to the edge of the wavelength range of BOSS) are evaluated over a $\pm$50 \AA\ window. We note that the width of the ``broad'' lines is driven by the maximum width seen in these lines across the entire sample, such as e.g., white dwarfs. And, although many lines that are typically narrow can be significantly broadened to widths of 100s of \kms\ ,thorough accretion or winds \citep[e.g.,][]{banzatti2019,sicilia-aguilar2020}, even in their most extreme cases they can easily fit within the 50 \AA\ window.

The flux in these windows is linearly reinterpolated onto uniformly-spaced, 128 element array with a pixel scale of $\sim$3.15 and $\sim$0.79 \AA\ respectively. Also, in order to normalize the flux in these windows, we transform it onto the log space.

The lines can significantly vary in strength. As such, to normalize their properties, we also take the log of the absolute value of EqW and AbW. However, while the AbW are always positive, EqW can be either positive or negative. In order to preserve this information, we add the sign of EqW to the output: the sign is set to +1 if EqW$>$0, -1 if EqW$<$0, and it is left at 0 if the line is undetected.

The model was constructed in TensorFlow, using a convolutional neural network (CNN) architecture. The data are passed through through 6 convolutional layers, with a convolutional kernel of 8 pixels; each layer breaks the data into an increasingly larger number of filters, from 8 to 32. Each convolutional layer is followed by a maxpooling layer to reduce the dimensionality of the data, and a $\tanh$ activation function. The outputs of the first six layers are flattened, and then are passed through 3 fully connected layers with 128, 256, and 128 neurons with ReLU activation function, after which they predict EqW, AbW, and EqW sign (which is also used as a classifier for the detection). In training, we used mean square error loss, masking it for EqW and AbW in sources where the line is undetected. 

Two models are trained, one for all of the broad lines, and one for all of the narrow lines (which refers solely to the adopted window size, and not the properties of the lines themselves). Doing so allows the CNN to better develop the ability to recognize the center of the window and to extract its properties. These models, however, can't be used as a general tool for extracting any lines not listed in Table \ref{tab:lines}, as the unfamiliar continuum can skew the weights. For example, the model trained on broad lines is generally able to reproduce EqW measurements of the narrow lines and vice versa (with the outputs scaled accordingly to their input pixel scales), but there are some systematic offsets, a large number of false positives and false negatives, as well as a significantly larger scatter than when the dedicated model is used.

\subsubsection{Testing}

The labelled data were split 80:20 into the train and test sets. We evaluate the performance of the models on the test set, and consider the line as ``detected'' in a given spectrum if the prediction for the EqW sign is $>$0.5 or $<-0.5$. Overall, across the sample the typical root mean squared in both EqW and AbW measurement is 0.09 dex For each transition, we examine the number of true positives (tp), false positives (fp), false negatives (fn), and evaluate both the precision (tp/(tp+fp)), and recall (tp/(tp+fn)). They are recorded in Table \ref{tab:lines}.

The model is performing extremely well with strong lines (e.g., H$\alpha$, with precision and recall of 98\% and 96\% respectively). In Li I, precision is relatively high, at 88\%, but its recall is somewhat lower, at 64\% - in these cases, the fp and fn sources are typically those that are borderline detections, with EqW$<$0.05 \AA. Some transitions do have quite low precision and recall. For instance, the He II 4685.7 \AA\ line has 8\% precision and 38\% recall, however, this transition is extremely rare, recorded for only 29 out of $\sim$3500 sources in the train set, of which only 8 are in the test sample. Thus, the ratios for this line are heavily affected by the small number statistics. When we evaluate the line on the full set of stars observed by BOSS (Section \ref{sec:eval}), when it is seen in absorption in cool stars, it is likely to be confused with other nearby features, but in stars with \teff$>$30,000 K, the absorption line measurements seem to be more robust, as do the rare cases when it is seen in emission, particularly for the ABYSS targets. 


\subsubsection{Evaluation} \label{sec:eval}

The resulting model is made available in \citet{lineforest}. We apply LineForest to all of the stellar sources observed by BOSS to-date, including all of the legacy data observed prior to SDSS-V. However, we note that the legacy SDSS data archive does not provide robust RV measurements as SDSS-V. This is mainly due to poor wavelength calibration, which was only identified at the beginning of the current iteration of the survey. As such, measurements of weak and narrow lines from legacy data spectra may be of lesser quality than in other datasets.

All of the lines were chosen to ensure they are detectable within the wavelength range of BOSS. LAMOST, however, has a narrower coverage, thus we only record the properties of the lines in the range of Pa11 to H$\epsilon$ inclusively.

We estimate the uncertainties in all of the measurements using a technique established in \citep{olney2020}, through computing several different realizations of a line by scattering the flux randomly by the reported uncertainties, passing each realization through the model, and averaging the predictions. In total, 100 iterations were made, and 1$\sigma$ errors were computed from 16, 50, and 84th percentiles.

The line measurements are reported only for the sources in which the line is detected confidently in at least 30 of these iterations, this suppresses the model from very noisy features. Sources with significant noise are also likely to have very uncertain RV measurements, thus making it difficult to identify weak lines in the first place. For example, 65\% of YSOs have $\sigma_{\rm RV}<$5 \kms. On the other hand, for YSOs with confident Li I detection, 75\% of them have $\sigma_{\rm RV}<$5 \kms, thus Li I measurements do indeed appear to be suppressed in noisy data.

\subsection{YSO Classifier \& Sample Definition}

The lines that have been measured with LineForest can be used to improve the identification of PMS stars. In order to utilize all of the features, we built a fully connected neural net using TensorFlow. We include \teff, \logg, all of the EqW and AbW, as well as six fluxes, from Gaia ($G$, $G_{BP}$, $G_{RP}$) and from 2MASS ($J$, $H$, $K$), for a total of 164 inputs. 

The network consists of three layers with 256, 512, and 1024 neurons connected using ReLU, a dropout layer with a rate of 0.5, distilling everything to a layer with a single output with a sigmoid activation function that returns the probability of a source being pre-main sequence. 

The initial training set consisted of all of the ABYSS sources that have been observed to-date, both with BOSS and with LAMOST, all of the SDSS-V BOSS sources, as well as a random subset of $\sim$1.4 million LAMOST field stars. 80\% of the sample was  used in training, and the remaining 20\% was used for testing. Since YSOs are relatively rare in the total training set, accuracy is an imperfect metric. Rather, we used a F1 score (the harmonic mean of the precision and recall) to evaluate the performance of the model.

While the targeting from ABYSS is rather comprehensive, it does include a significant fraction of more evolved field stars. Additionally, ABYSS may have missed some YSOs that have nonetheless have been serendipitously targeted by other programs. This mislabeling affects the quality of the model. To compensate for this, we reevaluate some of the labels.

First, we excluded ABYSS targets that are found at high galactic latitudes, such as $|b|>30^\circ$. We similarly excluded sources with RVs larger than what is typically found for the clustered sources at a given distance, sources that have \logg$<$3.4 or \logg$>$5.1, cool magnetically inactive stars with very low H$\alpha$ emission, all of the sources without Li I measurement and, finally, K dwarfs with insufficiently high Li I absorption. Also, we included all of the sources with H$\alpha$ emission that is consistent with originating from a protoplanetary disk, as well as K dwarfs with Li I absorption. By training a model using this sample, then we were able to examine the outputs, particularly for sources that the model identified as high confidence YSOs but were originally labelled as field stars. In that sample we interactively identified the sources that are consistent with being associated with young clusters and star forming regions that appeared to stand out as the overdensities in the plane of the sky, verifying that they are also clustered in proper motion and parallax phase space. These sources were then re-labelled as YSOs, and the model was retrained. This procedure was repeated a number of times until the number of ``discovered'' YSOs was negligibly small.

The final model was trained from a sample of 34,407 YSOs (down from 57,036 ABYSS targets). It achieves F1 score of 0.633 on the test sample, with precision of 0.752 and recall of 0.543 at a probability cut-off of 50\%. A higher probability cut-off improves precision, although we note that in that case, many of the ``false positives'' may actually be bona fide YSOs found in areas that are not strongly clustered in our sample. The bulk of the sources that are missing are found at higher \teff\ that are difficult to recover using traditional techniques as well, nonetheless we do note that, somewhat surprisingly, we are able to classify YSOs across the whole \teff\ range, and we are able to autonomously recover many of more distant populations that lack low mass stars in our sample (e.g., Cygnus X). The bulk of the information necessary for the classification is carried by just \teff, \logg, and EqW from H$\alpha$ and Li I, nonetheless other features improve the recovery of sources at these higher \teff. 

\begin{figure*}
\epsscale{1.1}
\plotone{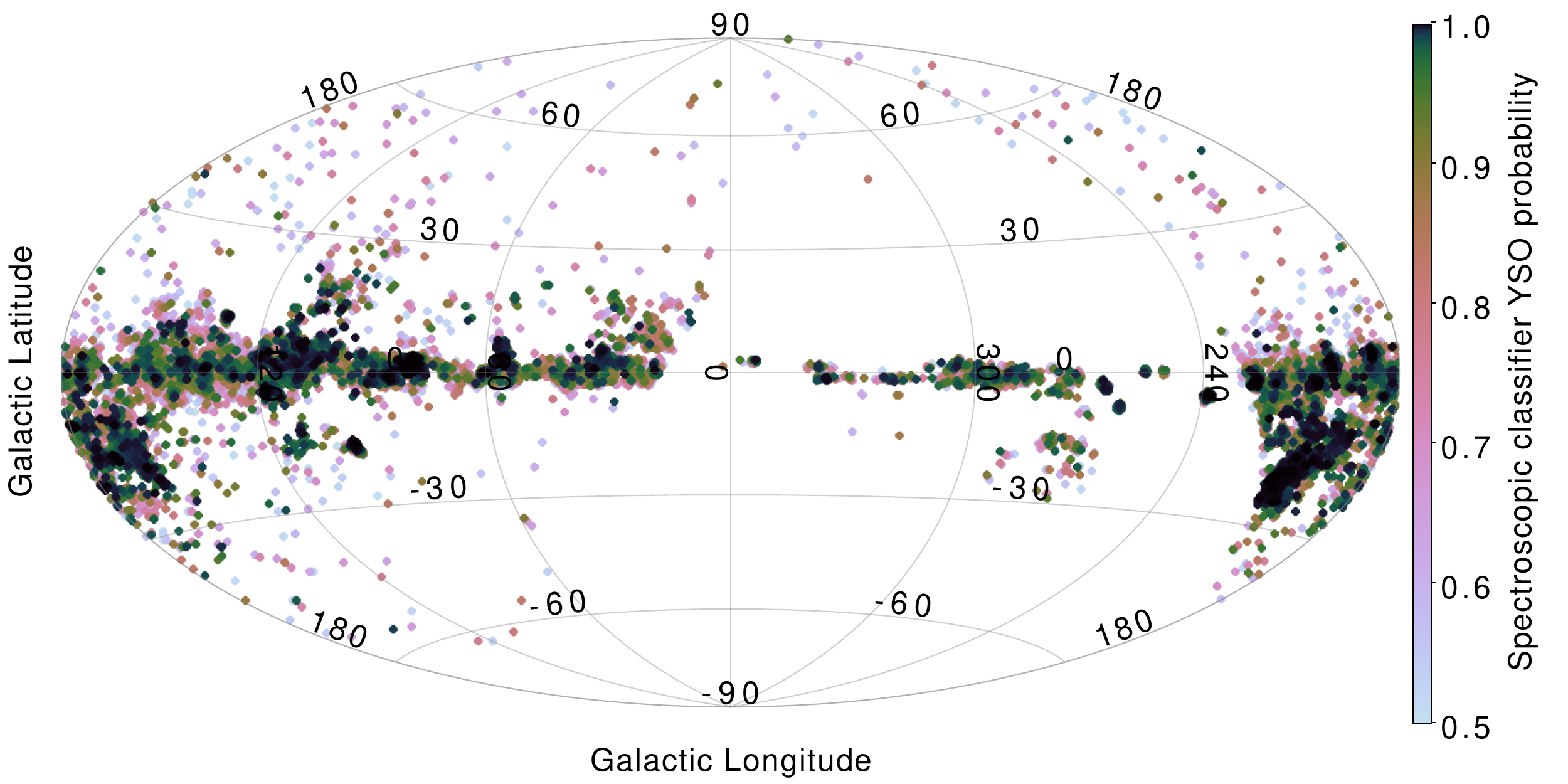}
\caption{On-sky distribution of identified YSO candidates with probability $>$0.5
\label{fig:sky}}
\end{figure*}

In total, we are able to identify, to-date, 17.9 K stars observed with SDSS-V BOSS as YSOs with probability $>$0.5, from which 11.1 K have a probability $>$0.8. Similarly, there are, 9371 stars in LAMOST DR8 with probability $>$0.5, of which 6105 stars have probability $>$0.8. Finally, there are 285 / 225 stars in the legacy SDSS data. Unsurprisingly, most of the sources are concentrated along the galactic plane and the Gould's belt, with only a few sources found at higher galactic latitudes (Figure \ref{fig:sky})

\subsection{Ages}

To examine the evolution of the sample (Section \ref{sec:li}), it is necessary first to determine the ages of the stars. For this purpose we use Sagitta \citep{mcbride2021}, a neural net that has two components. It first classifies sources into those that can be identified as pre-main sequence based on their photometry (G, BP, RP, J, H, K), parallax, and average extinction towards a given line of sight. Then, using the same input features, it estimates photometric ages for the individual PMS stars. It has been trained on the average ages within a subcluster in which a given star is found, and at the moment this pipeline offers the most stable performance with respect to \teff\ (i.e., producing self-consistent ages for G, K, and M-type stars in a given population), in comparison to the more traditional techniques that rely on the theoretical isochrones \citep{kounkel2023a}.

\begin{figure}
\epsscale{1.1}
\plotone{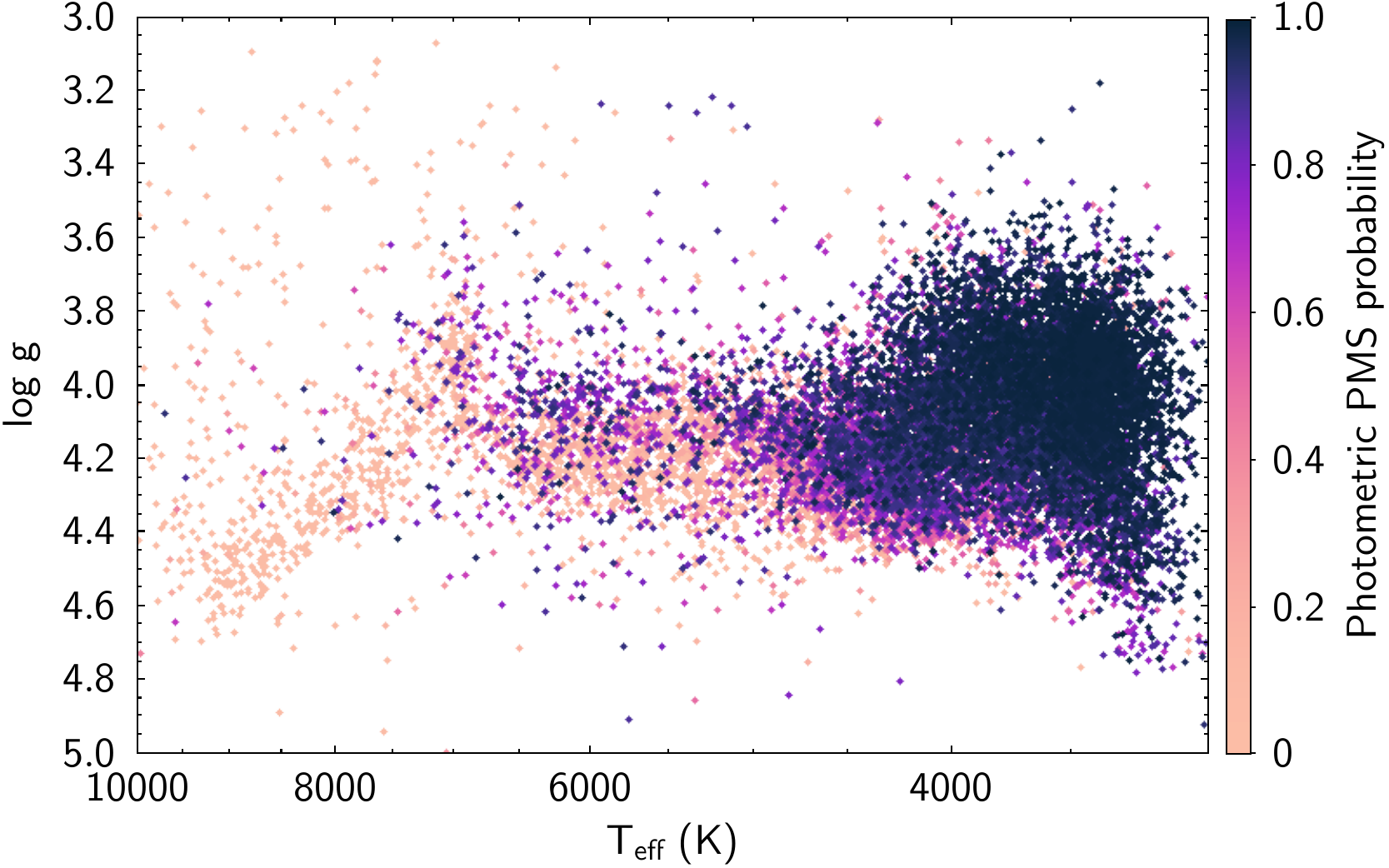}
\plotone{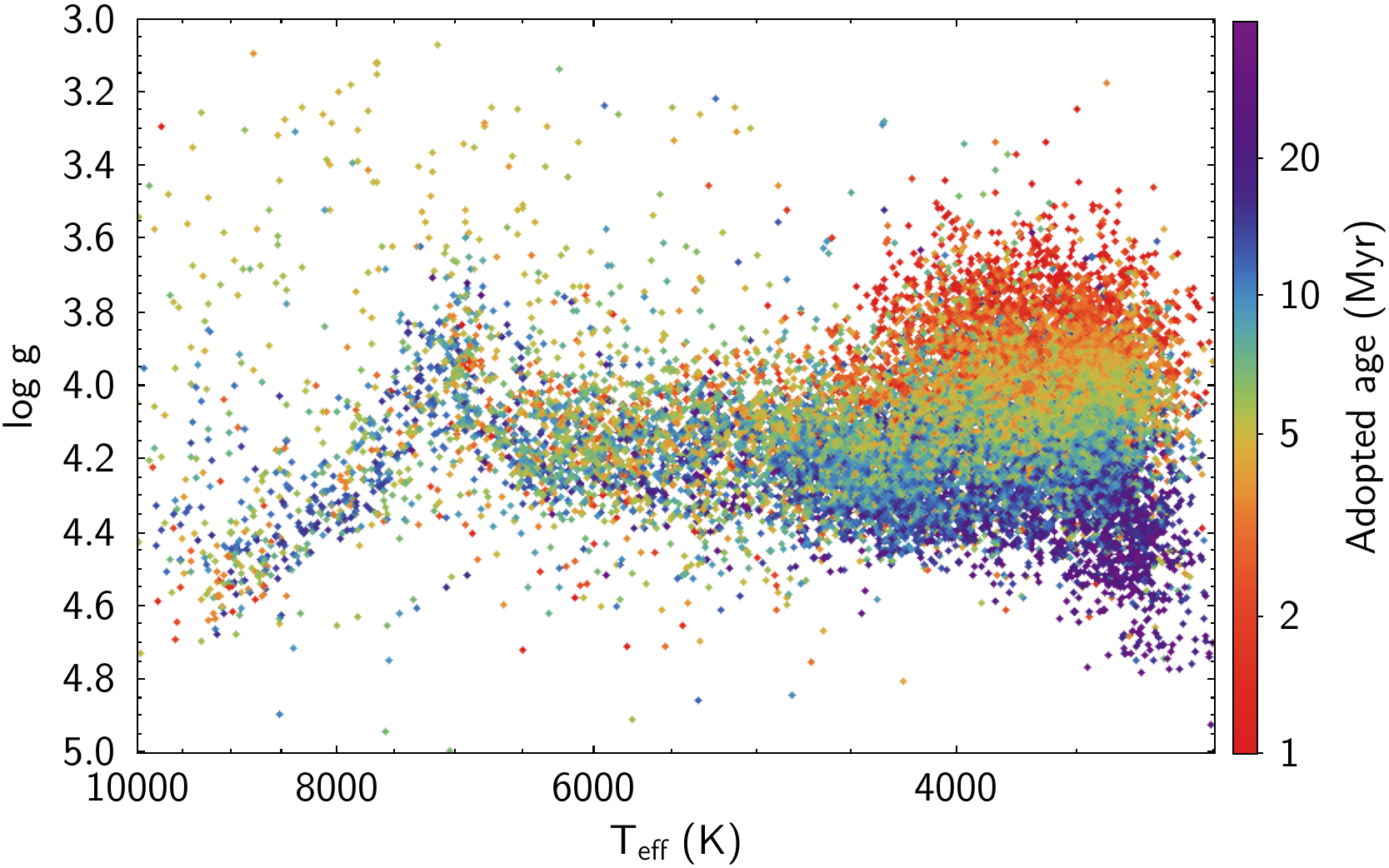}
\caption{Top: Kiel diagram of the sample, color coded by the photometrically derived probability of a star being PMS using Sagitta \citep{mcbride2021}. Bottom: Adopted ages for the stars in the sample; for the sources with PMS prob.$<$0.5, the assumed age is that of a closest high probability neighbor.
\label{fig:pmsage}}
\end{figure}

Nonetheless, Sagitta is only reliable for the stars that are classified as PMS. Higher mass stars quickly reach the main sequence, and as such, their photometry stops being a reliable indicator of their ages in comparison to the lower mass stars of the same age. For this reason, in the total sample of stars with spectroscopic YSO probability $>$0.5, there are a number of sources ages of which has to be derived through other means to avoid biases in the overall distribution (Figure \ref{fig:pmsage}).

To fill in this gap we only retain ages for those stars with Sagitta-derived PMS probability of $>$0.5. We then calculate the 3-d distance from a lower probability to a closest higher probability source. We then assign the age of that closest high probability source if the separation is $<$30 pc. We have experimented with different cut-offs to observe the effects on the analysis presented later in the paper, to achieve an appropriate trade-off in terms of completeness at higher \teff\ while maintaining robustness of the sample. Although this separation is considerable, it does incorporate within it a typical uncertainty in the parallax, and it is also permissive for the inclusion of the more diffuse populations. On the other hand, in denser regions, the closest neighbor is usually well under that limit.

Such substitution has been done for 26\% of the full sample, with the bulk of it dominated by hotter stars. Among sources with \teff$>$5000 K, 74\% of sources have their ages substituted, on the other hand, this is the case for only 12\% of cooler stars, most of which have age $>$10 Myr.

\begin{figure}
\epsscale{1.1}
\plotone{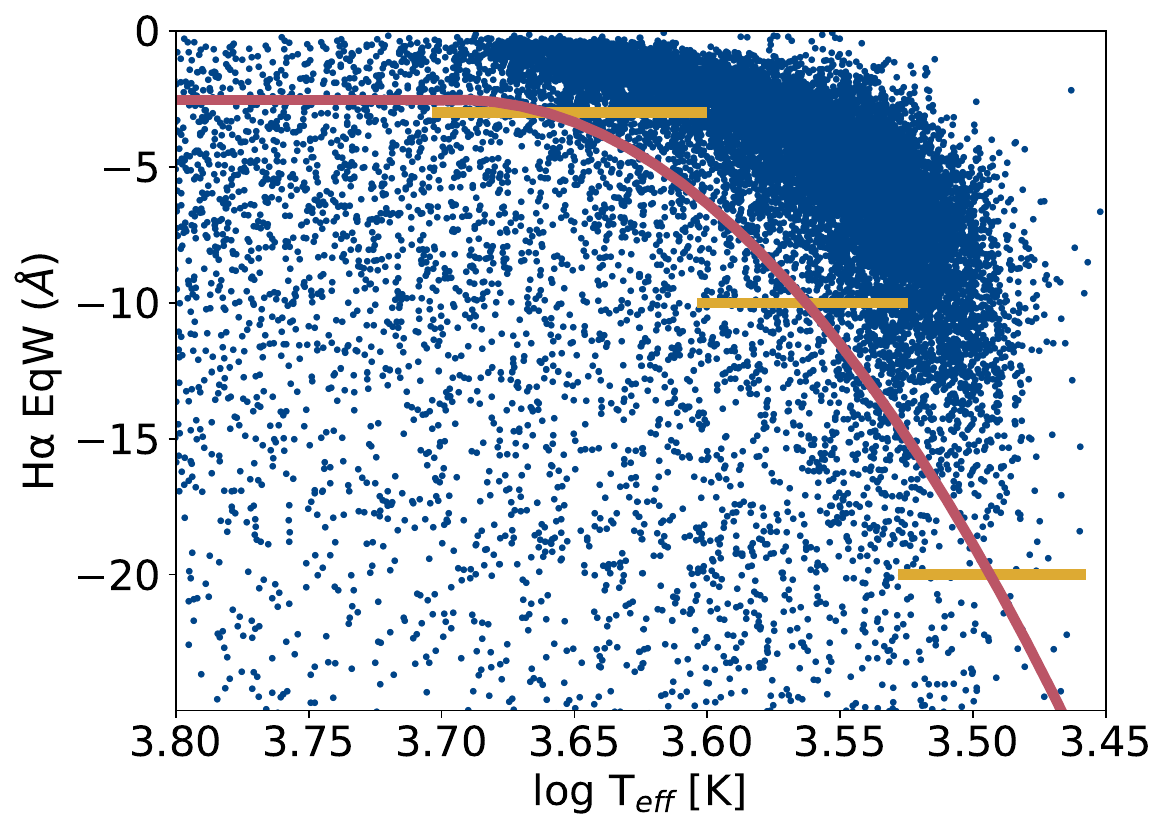}
\caption{Distribution of H$\alpha$ as a function of temperature for the YSOs in our sample. The red line shows the adopted separation between CTTSs and WTTSs. The yellow lines are the ranges for the selection from \citet{white2003}.
\label{fig:ctts}}
\end{figure}

\begin{figure*}
\epsscale{1.1}
\plottwo{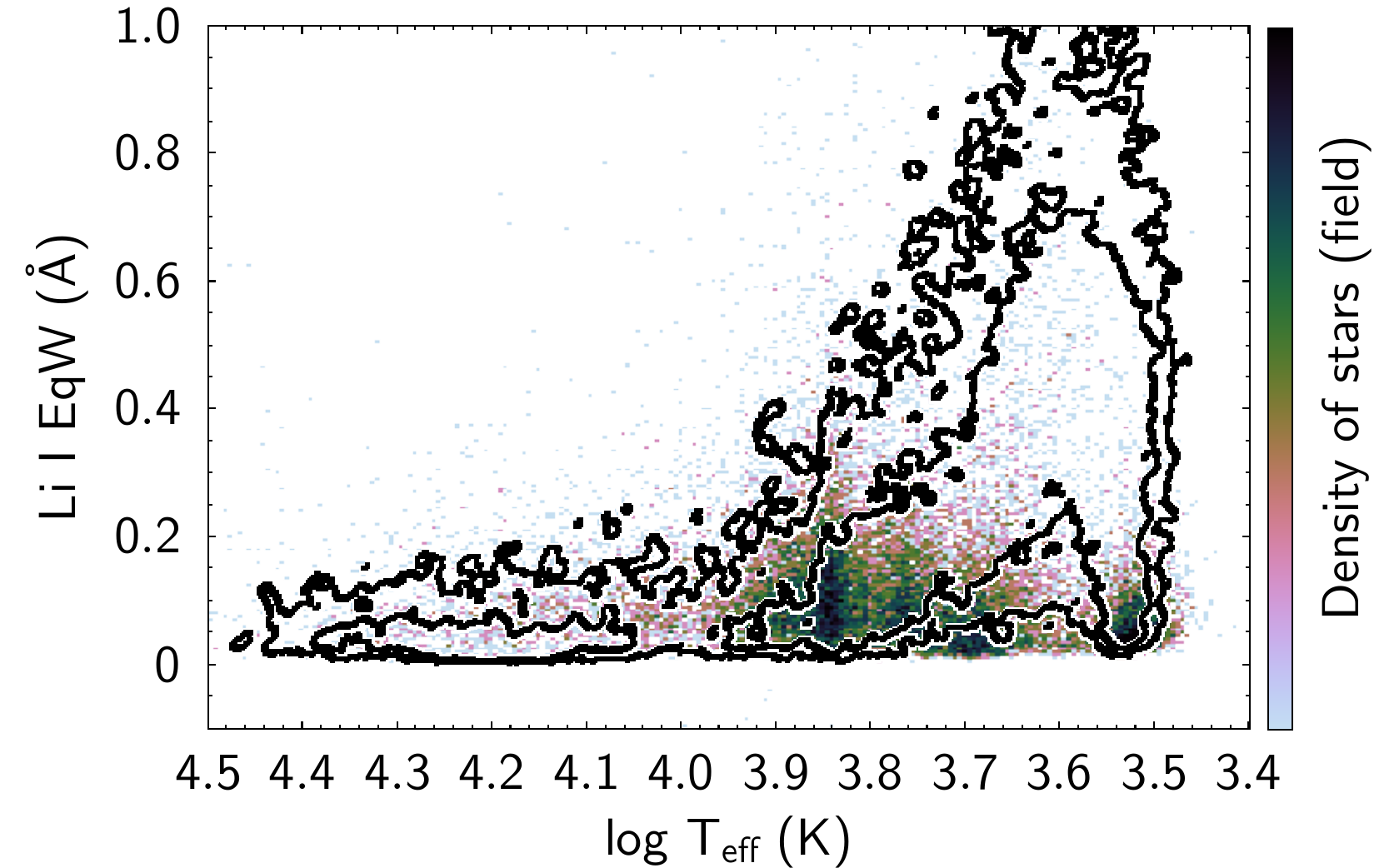}{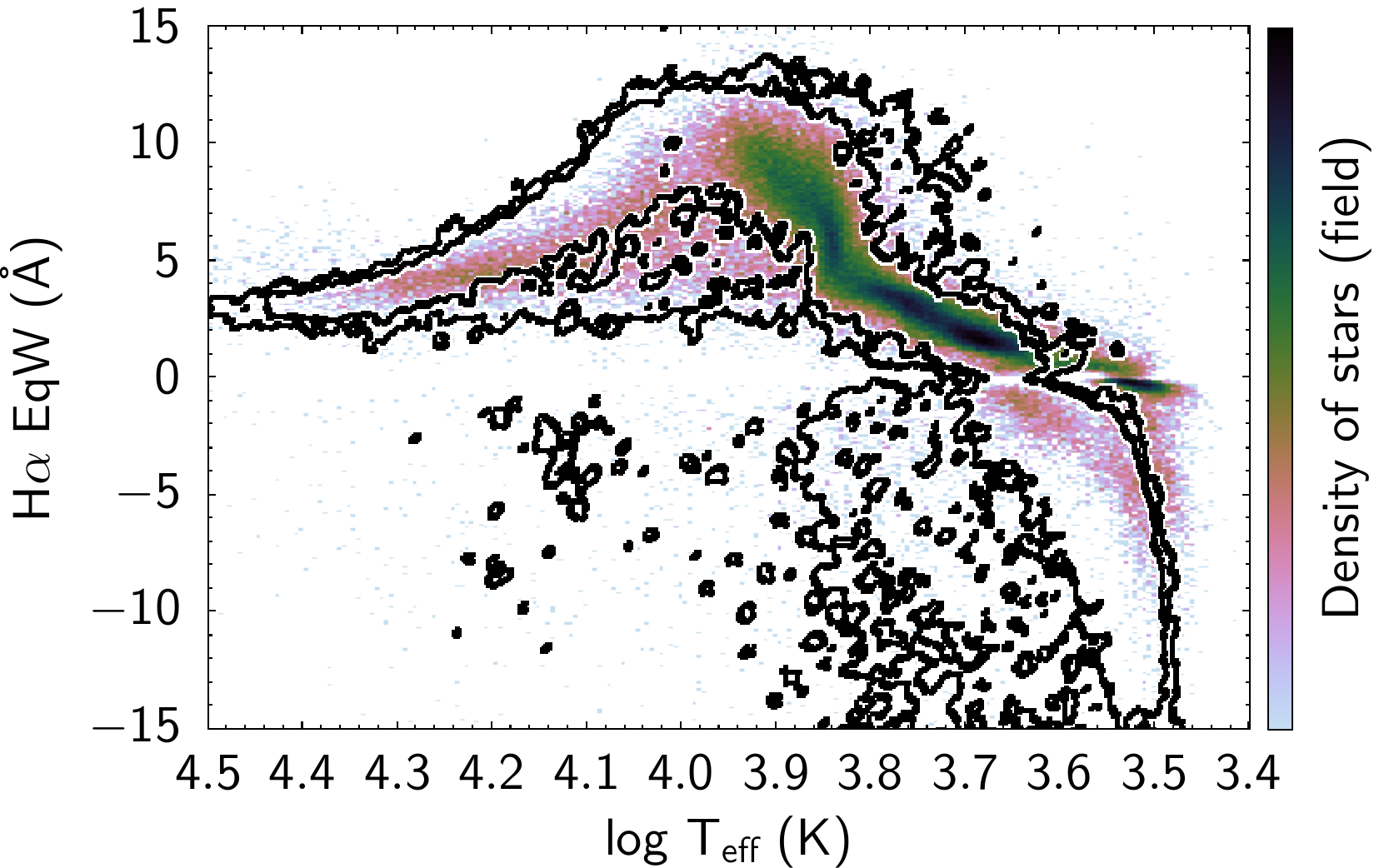}
\plottwo{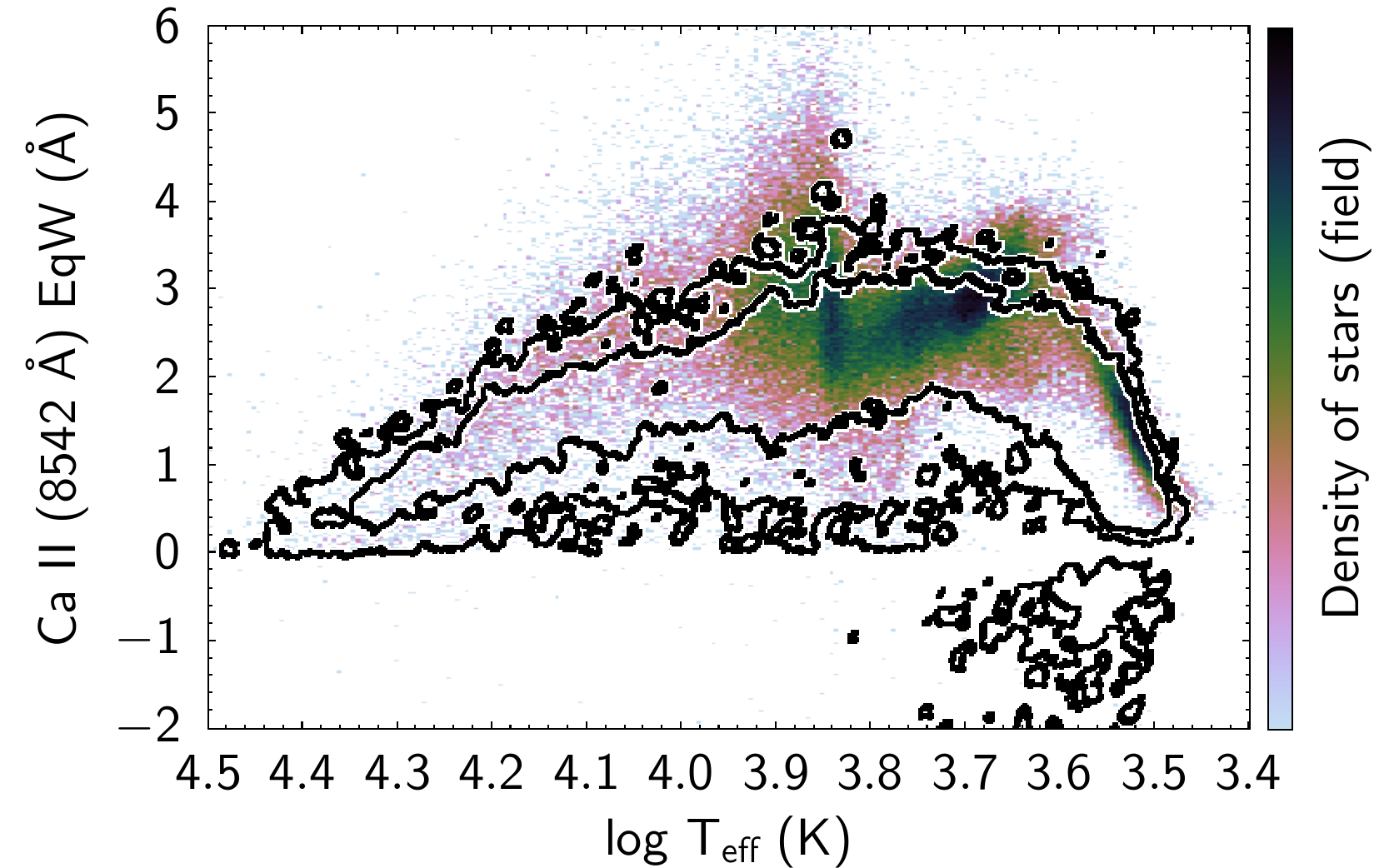}{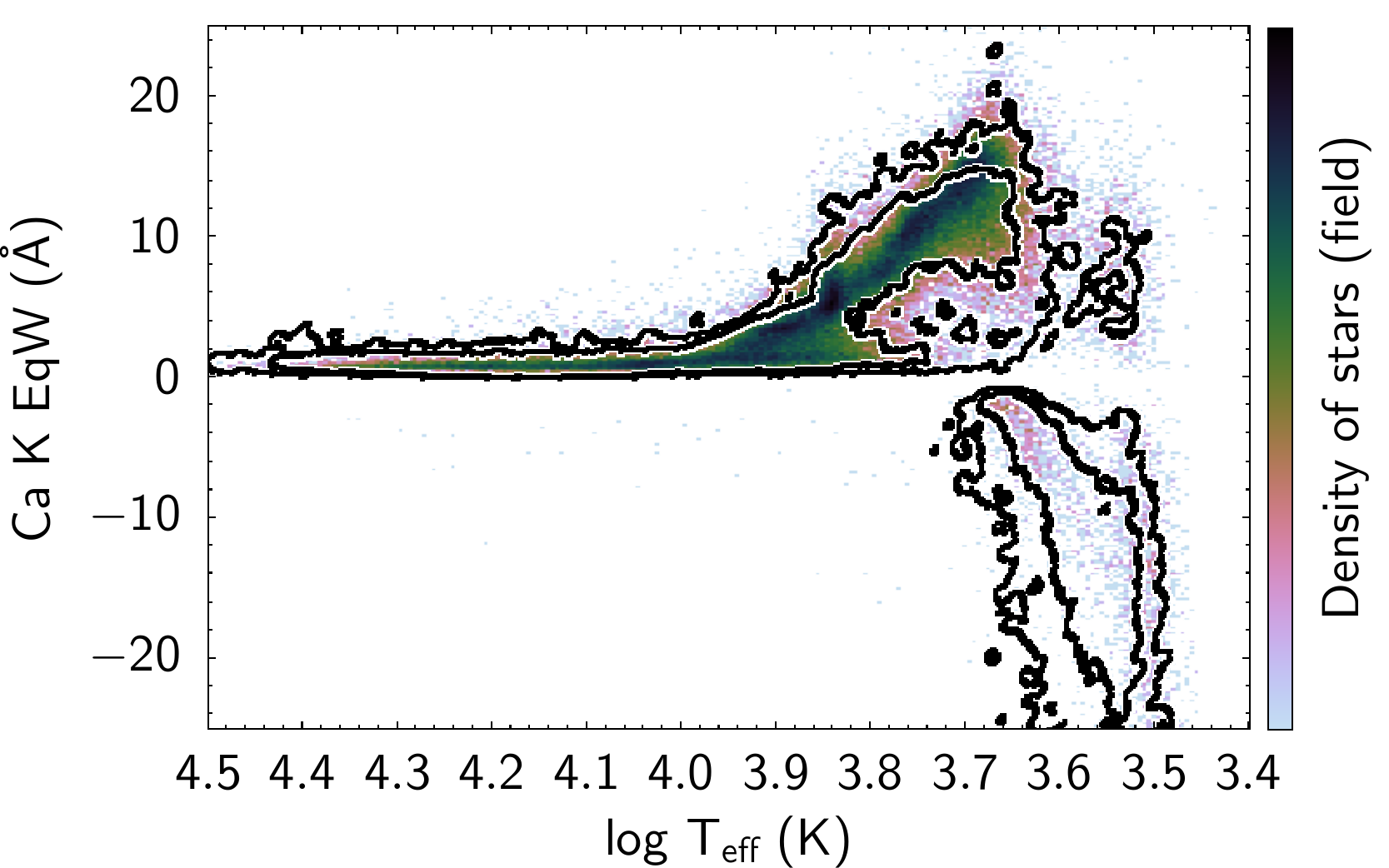}
\plottwo{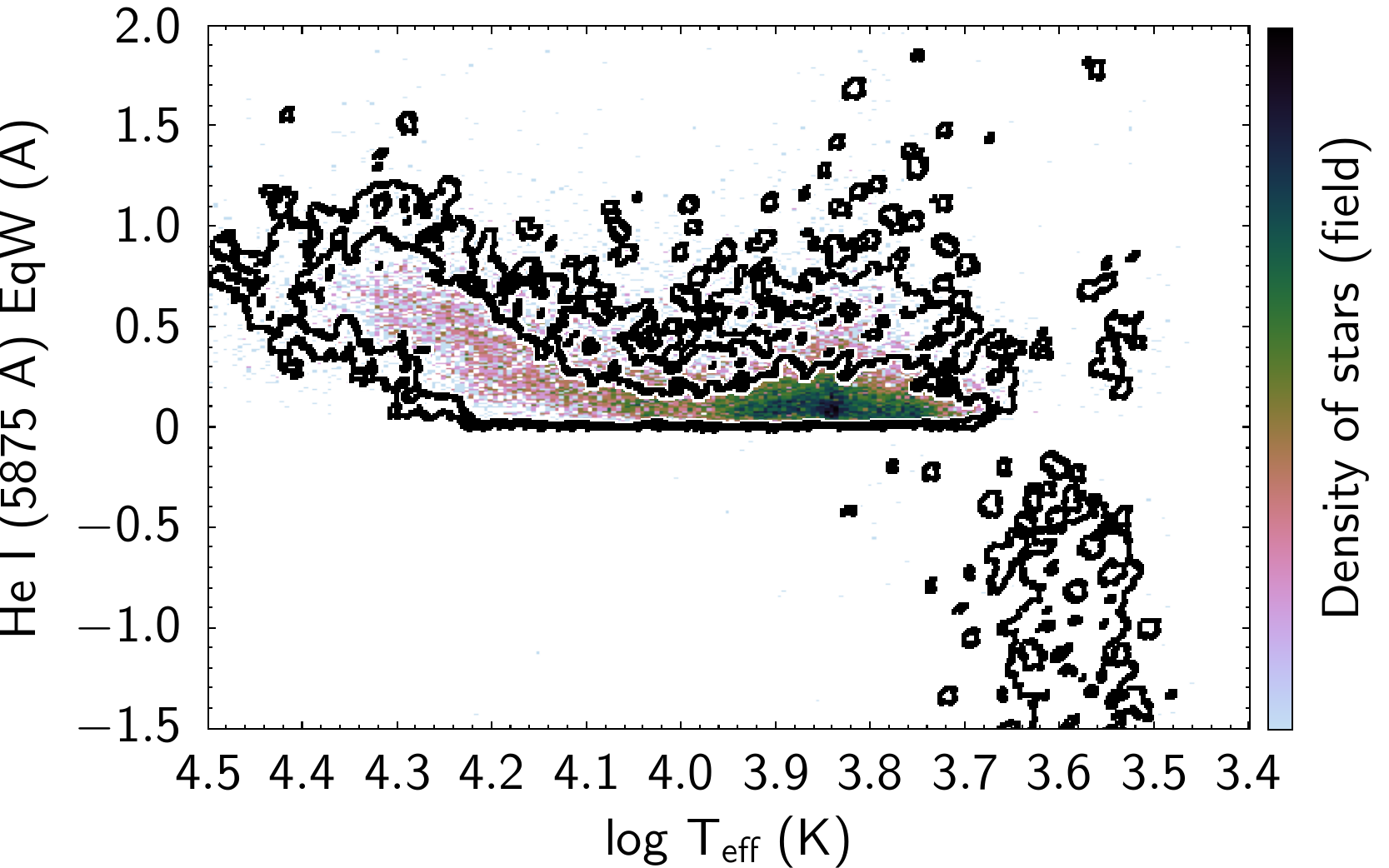}{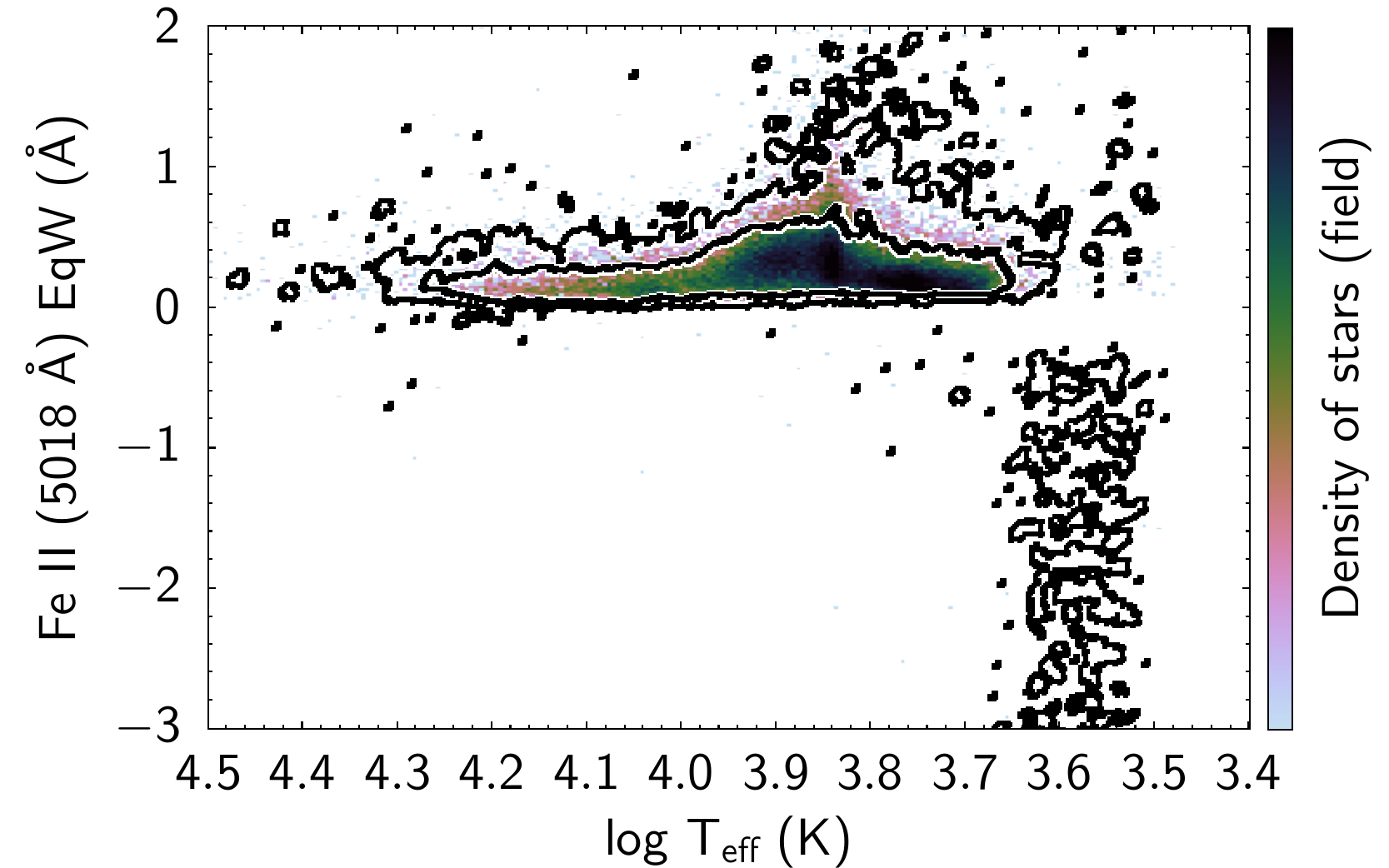}
\plottwo{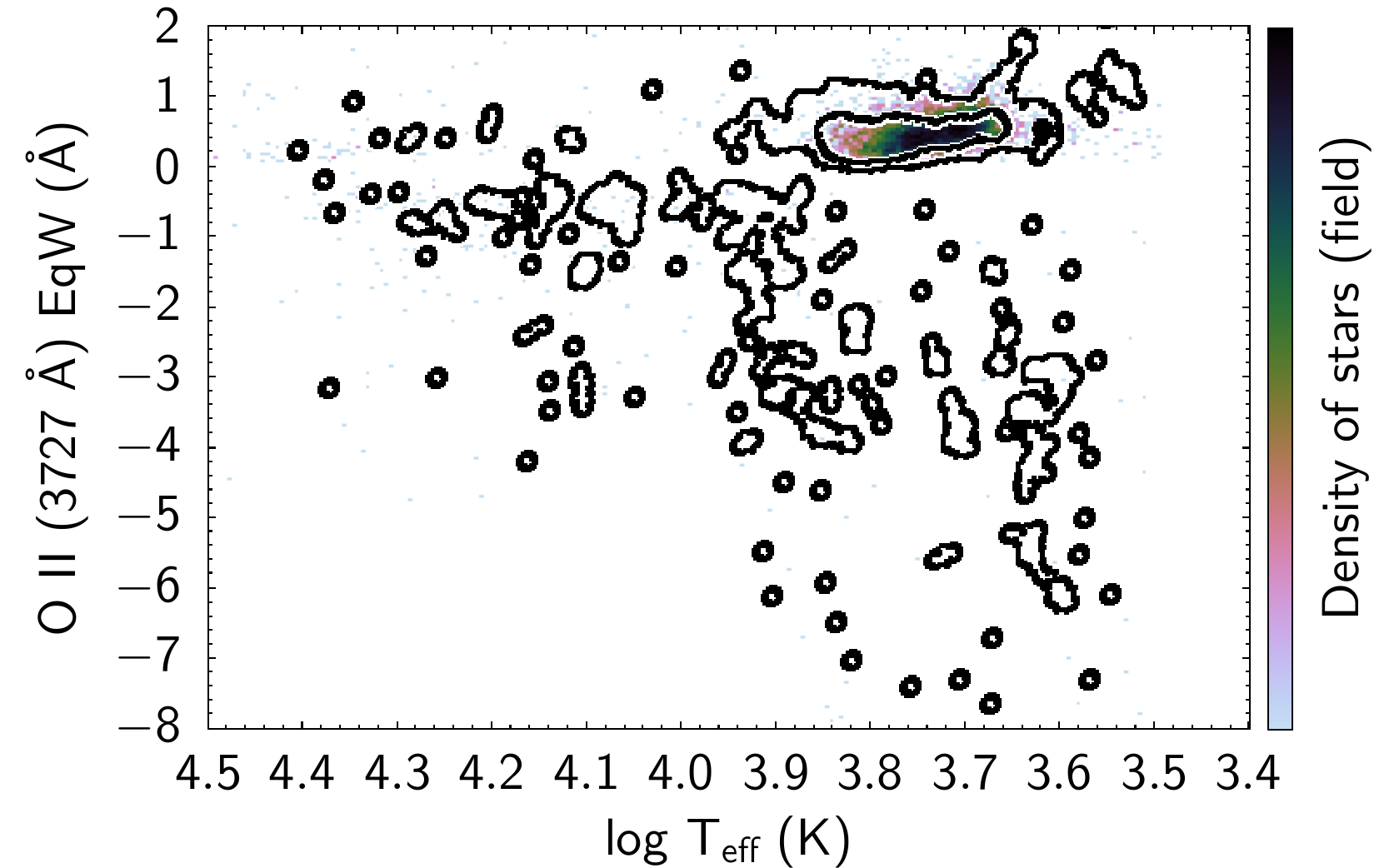}{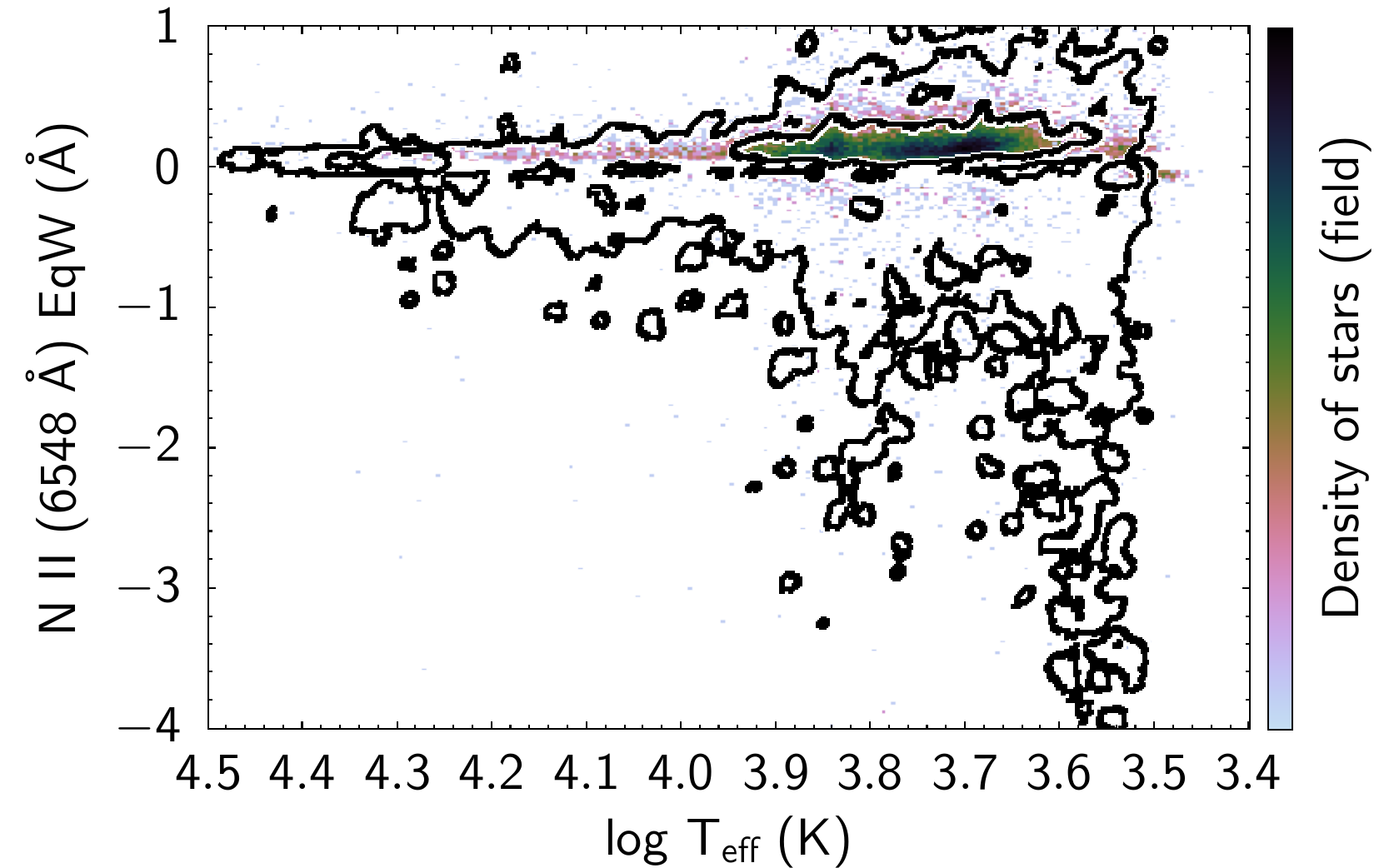}
\caption{EqW measurements in BOSS spectra for some of the lines used in this study. The colored background (color coded by the density of points) represent field stars of different evolutionary stages, black contours are the sources identified as YSOs.
\label{fig:lines}}
\end{figure*}

\subsection{Identification of CTTSs}

\citet{white2003} have developed a traditional method to select CTTSs based on H$\alpha$ emission strength as a function of spectral type, employing 4 flat cuts. The simplicity of its implementation becomes a downside when dealing with a very large sample, as the transition between these cuts become very pronounced. \citet{briceno2019} have improved on this approach by developing a much more continuous selection, but their approach relies strongly on spectral types of the stars. Commonly used transformations between spectral type and \teff\ exist \citep[e.g.,][]{pecaut2013}, but since we do have measurements of \teff\ available directly, it is much more optimal to use a selection criterion that does not rely on such a conversion.

After examining the distribution of H$\alpha$ EqW as a function of \teff, we notice a clear concentration of sources with weak emission lines (i.e, WTTSs). On the other hand, among those sources with stronger emission lines (i.e, CTTSs), there is a much more significant scatter (Figure \ref{fig:ctts}). Using this overdensity of WTTSs, we define a cut
\begin{equation}
\begin{split}
    \log T_{\rm eff}<3.7: H\alpha<-436.95\log T_{\rm eff}^2+\\3227.7\log T_{\rm eff}-5963.2 \\
    \log T_{\rm eff}>3.7:         H\alpha<-2.55
\end{split}
\end{equation}
\noindent to select out CTTSs, which we then compare against \citet{white2003} with the transformation from \citet{pecaut2013} as a guide.

\section{Results}

We present LineForest measurements for 23,903 stars identified as YSO candidates in LAMOST spectra with probability $>$0.1 in Table \ref{tab:lamost}. Outputs for all of the BOSS spectra (including non-YSOs), will be incorporated in the subsequent SDSS data releases.

\begin{deluxetable}{ccl}[!ht]
\tablecaption{LAMOST PMS candidates
\label{tab:lamost}}
\tabletypesize{\scriptsize}
\tablewidth{\linewidth}
\tablehead{
 \colhead{Column} &
 \colhead{Unit} &
 \colhead{Description}
 }
\startdata
obsid & & LAMOST unique identifier \\
RA & deg & Right ascention in J2000 \\
Dec & deg & Declination in J2000 \\
log \teff & [K] & Effective temperature from Sizemore et al. \\
$\sigma$ log \teff & [K] & uncertainty in log \teff \\
log g &  & Surface gravity from Sizemore et al. \\
$\sigma$ log g & & Uncertainty in log g \\
RV & \kms & Radial velocity from Sizemore et al. \\
$\sigma$ RV & \kms & Uncertainty in RV \\
PMS & & Probability from the spectroscopic classifier of the \\
    & &  object being a PMS candidate, reported for PMS$>$0.1 \\
\_eqw & \AA & Equivalent width, all the lines in the sample\\
$\sigma$ \_eqw & \AA & Uncertainty in \_eqw \\
\_abw & \AA & Absolute width, all the lines in the sample\\
$\sigma$ \_abw & \AA & Uncertainty in \_abw \\
\enddata
\end{deluxetable}

\subsection{General line properties}

We examine the overall differences in the measured properties of the lines between stars identified as young and the significantly more evolved field stars (Figure \ref{fig:lines}).

As expected, Li I shows an obvious signature of evolution. In our sample, we observe how the YSO candidate distribution clearly peaks near 0.45 \AA\ at \teff$\sim$4000 K, while only a trace amount of Li I is observed at this \teff\ in the field stars, largely within the errors. While there is some variance in its distribution with \teff, it is not significant. On the other hand, in YSOs Li I abundance significantly decreases towards the cooler end, since these stars become fully convective, resulting in a higher internal \teff\ that is available for Li I depletion. Similarly, towards the hotter end, the internal temperature of the star itself increases significantly, even though the convective envelope shrinks. The overall shape of Li I distribution in young stars is consistent to what has been observed in the past \citep[e.g.,][]{jeffries2014,jeffries2023}. We discuss evolution of Li I in greater detail in Section \ref{sec:li}.

H$\alpha$ is another notable line. It is commonly seen in emission in cool stars, and in absorption at stars with \teff$>$6500 K, reaching a maximum EqW at $\sim$10,000 K. There is a difference in the EqW in early-type stars, with younger stars typically having stronger absorption lines; this difference is most likely driven by different \logg\ distribution (Sizemore L. et al. submitted). Young high mass stars are expected to be on the main sequence, on the other hand, since high mass stars have short lifetimes, older stars rapidly evolve away from it.Among late type stars the difference is much more significant. 
This is apparent not just in CTTSs (which show H$\alpha$ emission lines far in excess of what is observed in the field), but in WTTSs as well. Evolved M dwarfs in the field can be divided into inactive (with H$\alpha$ EqW$\sim$0 \AA), and those that are magnetically active, where H$\alpha$ has stronger emission \citep{newton2017}. Both are present in SDSS-V sample. Active M dwarfs are usually considered to be relatively young, with ages younger than a few Gyr \citep{west2008}. Here we are able to observe a significant evolution between these active M dwarfs and WTTSs, with H$\alpha$ emission lines being 1.5-2 times stronger in WTTSs than in somewhat older field stars. This is an indication of the magnetic activity decreasing in strength as stars age.

A line behavior similar to that of H$\alpha$ is observed in other H lines as well, both in the Balmer and in the Paschen series, though, higher energy lines are less likely to be seen in cooler stars. Similar trends can also be observed in Ca lines as well -- they also show evolution of \logg\ among early type stars (although whether YSOs have stronger or weaker lines usually depends on the precise type of transition), and they also separate late type stars based on their age. The Ca triplet usually consists of narrow lines that are seen in emission only in some CTTSs. And although they are primarily seen in absorption, low mass YSOs tend to have weaker Ca triplet lines. This effect is likely to be also primarily driven by the magnetic activity filling the lines (but not to the degree of creating emission lines), as these lines are magnetically sensitive, and the excess flux in them typically correlates with emission in Ca H \& K \citep{martin2017}. However, veiling could also have some influence on these lines.

Some differences are observed between the Ca triplet, and the Ca H \& K lines. The latter two tend to have very broad absorption lines in which a narrow emission component may emerge. This results in a discontinuity at EqW=0 \AA\, as the pipeline transitions from measuring the full absorption line to only measuring its emission portion. Low mass YSOs are commonly seen with Ca H \& K in emission, this is only rarely the case in the field stars \citep[in which case they are most likely to be younger than a few 100 Myr,][]{cunningham2020}.

CTTSs produce emission lines from many different transitions. Often (e.g., in He I or Fe II lines), this occurs in cool stars, with \teff$<$4500 K presenting very strong H$\alpha$ EqW$<$-60 \AA. Some transitions (e.g., O II or N II) can produce emission lines across all \teff. In early type stars this may be a more reliable indicator of accretion than H$\alpha$.

\subsection{Hydrogen decrements in CTTSs}

\begin{figure*}
\epsscale{1.1}
\plottwo{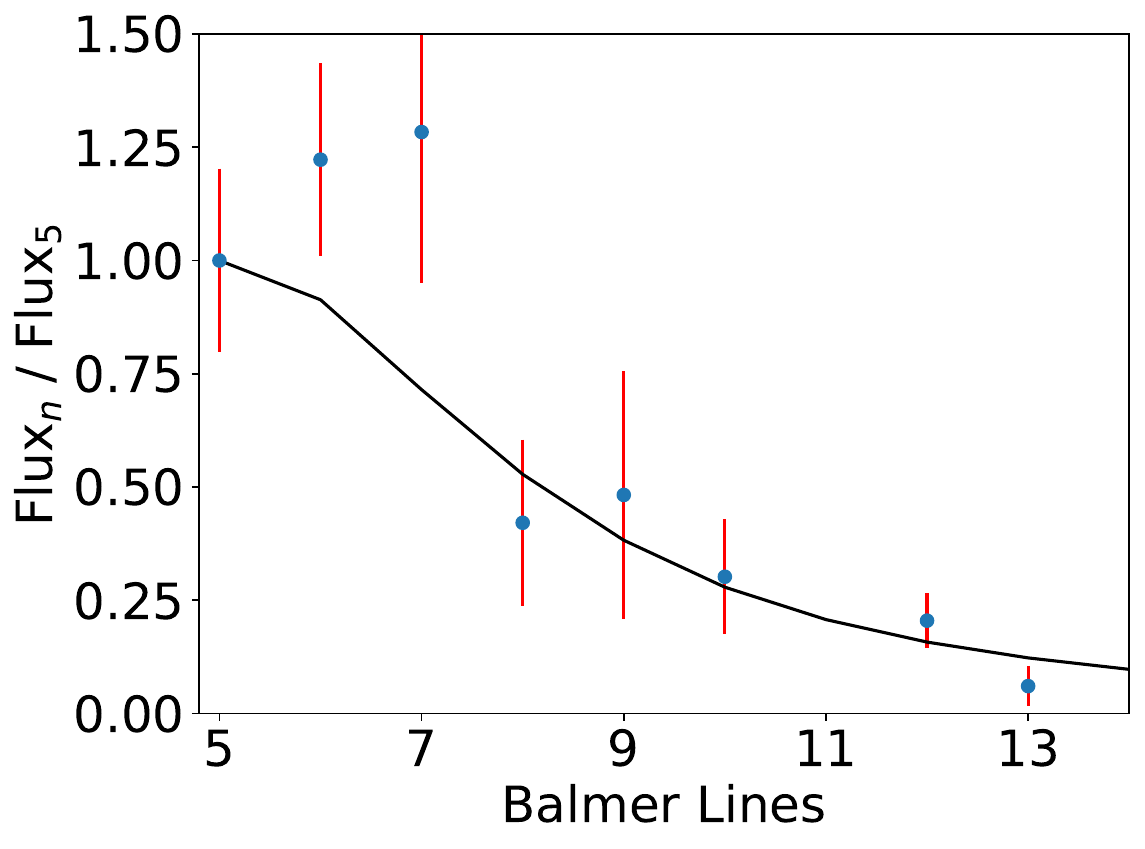}{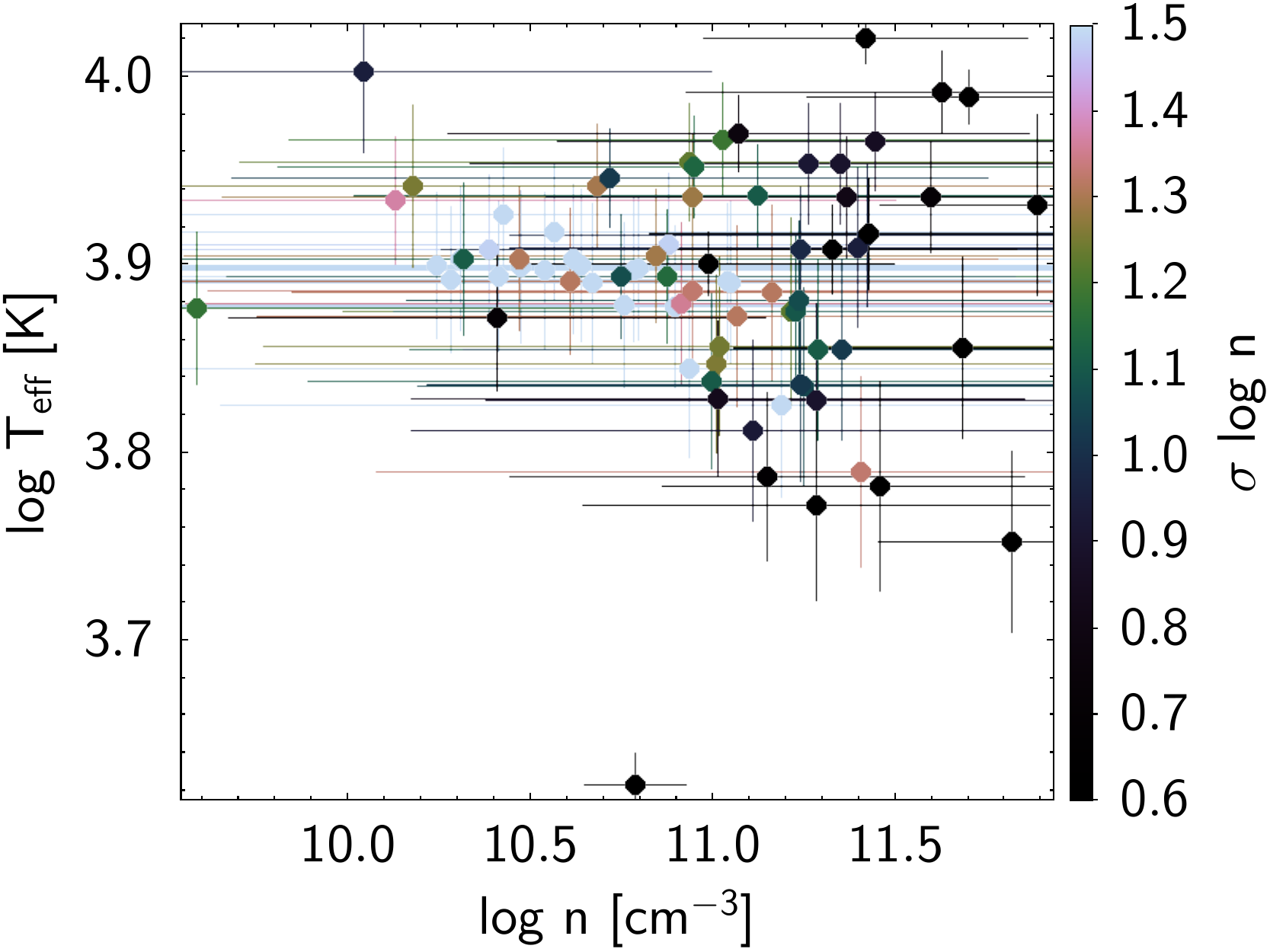}
\caption{Left: An example of the decrement of EqW of H lines in the Balmer series, normalized by EqW of H$\gamma$ ($n_u=5$). The black line shows the best fitted model from \citet{kwan2011}. H$\epsilon$ is likely contaminated from Ca H. Right: Distribution of the derived Temp and log n models for the sources with cleanly defined decrements in Balmer lines. Points are color coded by the uncertainty in log n, note that most confident measurements tend to have log n$>$11.
\label{fig:decrement}}
\end{figure*}

While CTTSs are defined using the strength of H$\alpha$, it is not uncommon to see emission lines in other H lines as well. Of 6444 CTTSs, 3085 show emission in H$\beta$, 1966 in H$\gamma$, 1215 in H$\delta$, 1021 in H$\epsilon$, etc.

When multiple lines are available, since each of them has different excitation temperature, it is possible to probe the physical conditions such as temperature (Temp) and density (log n) of the emitting gas (i.e, the accretion stream). This was previously done in \citet{campbell2023} using Brackett lines observed with APOGEE, through comparing the ratios of equivalent widths of these lines to the models from \citet{kwan2011}.

We attempt to replicate this experiment using the Balmer lines (Figure \ref{fig:decrement}). In CTTSs we isolate the H lines in all of the sources that appear to be in emission. We then compared them to the models from \citet{kwan2011}. Since some of the H lines are likely to be optically thick (and thus their EqW would not scale accurately with the abundances), we examined different subsets of these lines, normalizing all EqWs by the strongest line in the subset.

Because of this we discarded H$\alpha$ and H$\beta$ from the analysis; the higher order lines on the other hand produced a more self-consistent fit. Thus, we limited the sample only to the sources that have H$\gamma$ and at least two higher energy lines to construct a decrement. Similarly, we attempted to examine the decrements in the Paschen series, but while results were broadly consistent to the Balmer series fit, there was significant uncertainty in the fit, thus they were excluded from the analysis.

Through the model comparison, we identified the Temp and log n of the gas. We note that we have not corrected EqWs for veiling or extinction, thus the presented results should be considered preliminary, and done only for the sources with a clearly defined decrement. An estimation of these parameters will be done in the subsequent papers in the series. 

We find that Temp of the accretion shock is typically $\sim$6000--10,000 K, and that density is $\sim10^{11-11.5}$ cm$^{-3}$ for the sources with the most confident measurements. There is no strong correlation with the properties of the star itself, such as \teff, \logg, age, or any other. 


For comparison, \citet{campbell2023} find a much larger range in Temp, from 4000K to 16,000 K (with the sources with hotter Temp preferentially being Be stars) and density ranging from $\sim10^{12}$ cm$^{-3}$ at 5000 K, to $\sim10^{11}$ cm$^{-3}$ at 12,500 K. Over the range of the overlapping Temp, there is a relative agreement in the derived log n.

\section{Discussion: Li I evolution with age} \label{sec:li}

\begin{figure}
\epsscale{1.2}
\plotone{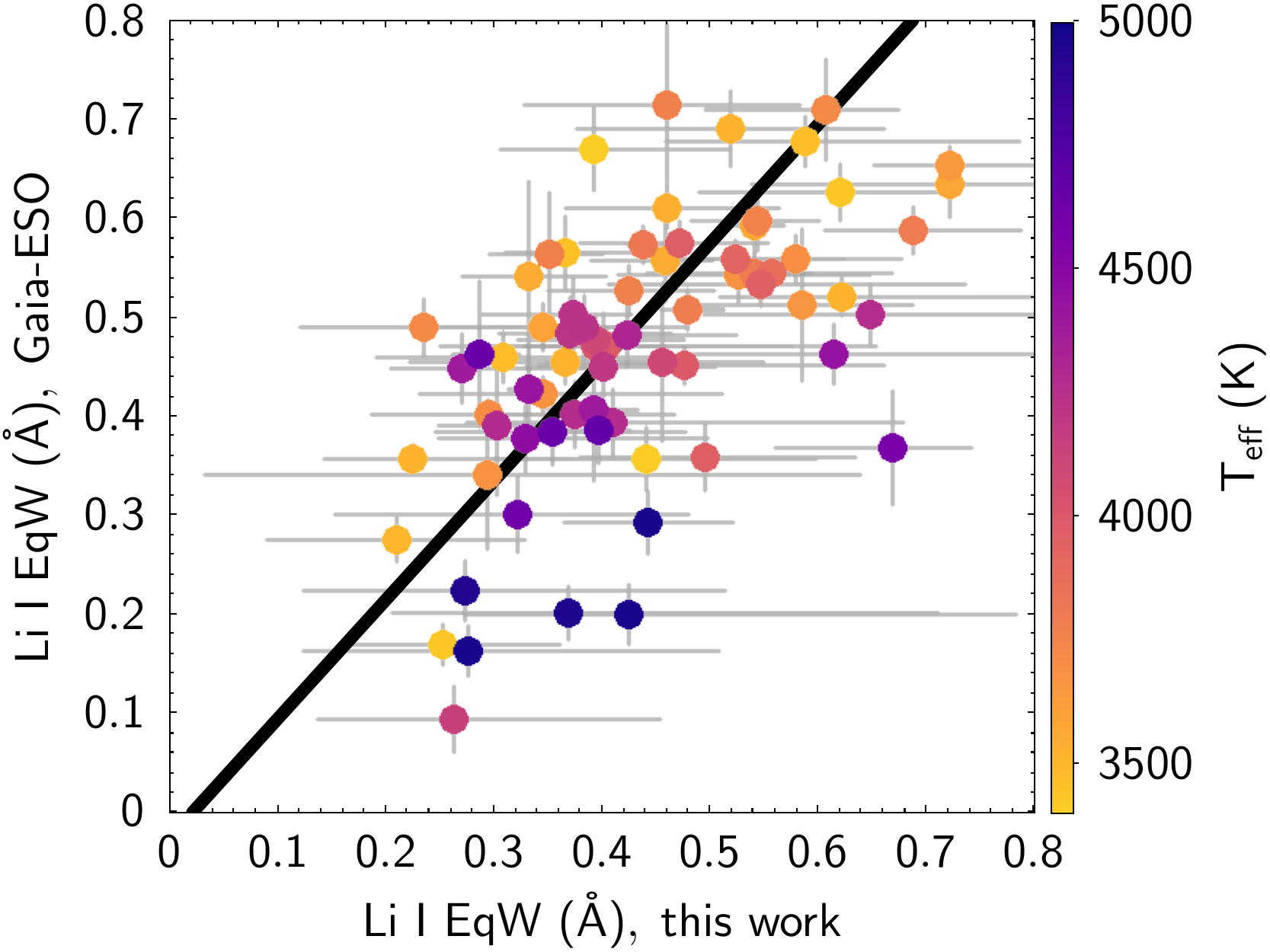}
\caption{Comparison of Li I measurements reported here to those from Gaia-ESO spectra from \citet{binks2022}, color-coded by \teff\ of the star. The black line shows the best fit between the two.
\label{fig:eso}}
\end{figure}

\begin{figure*}
\epsscale{1.1}
\plotone{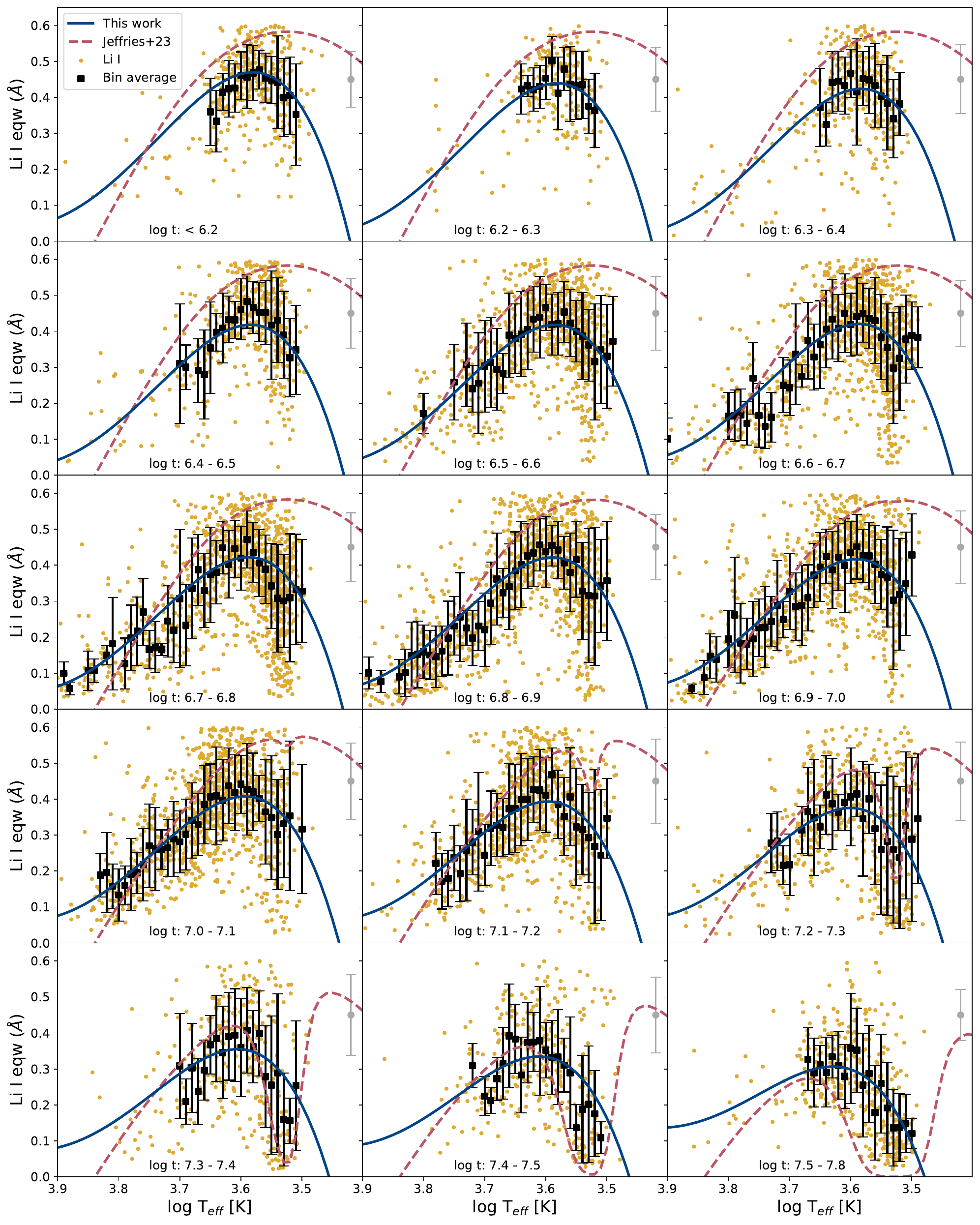}
\caption{Li I distribution as a function of \teff\ across different age bins. Yellow dots show the raw data, black squares show the median Li I EqW in 0.1 dex \teff\ bins, and the black error bars show the 16-84 percentile scatter. The grey errorbar shows the typical uncertainty in the individual measurements. The red dashed line shows the relationship of Li I as a function of \teff\ and age from \citet{jeffries2023}, the blue solid line shows the polynomial fit presented in Equation \ref{eqw:poly}
\label{fig:liall}}
\end{figure*}

To assess the reliability of the measurements, in Figure \ref{fig:eso} we compare our measurements of Li I with the measurements reported by \citet{binks2022} from Gaia-ESO spectra. Gaia-ESO spectra have significantly higher resolution ($R\sim$15,000) than BOSS or LAMOST spectra, making these measurements significantly more robust. Indeed, the comparison does show significant scatter, but this scatter is mostly reproduced by our reported uncertainties. Furthermore, there is an excellent correlation between the two sets of measurements, particularly at \teff$\leq$4500 K, although Gaia-ESO measurements appear to be systematically larger by $\sim$0.05 \AA\, likely due to the differences in the continuum definition.

In young stars Li I is known to precipitously deplete as they get older. Recently, \citet{jeffries2023} have developed an empirical model of Li I depletion using Gaia-ESO measurements of 52 open clusters with age ranges from 2 Myr to 6 Gyr. While it serves as a good picture of the overall global evolution of Li I, the sampling at the younger age ranges is somewhat poor, and no attempt was made to distinguish between clusters younger than 10 Myr. Since our sample has more than an order of magnitude more star younger than 25 Myr, we reexamine the trend at these age ranges.

To do this, we separate the sources 0.1 dex bins both in age and in \teff, and we find the median, as well as 16th and 84th percentile of the distribution to check the scatter (Figure \ref{fig:liall}). We also find the typical uncertainty in Li I measurement of the sources in the bin (to evaluate the significance of the scatter), as well as the weighted average error for the median itself. Since the external comparison does show our uncertainties as robust, even with considerable individual errors, a sufficiently large statistical sample allows improving on finding the mean of the distribution.

The relation from \citet{jeffries2023} appears to be somewhat inconsistent in describing our Li I measurements of these young stars. In part it is due to the systematics in the assumptions of age and \teff, as well as the aforementioned systematic offsets in our determination of Li I, but also it can be attributed to the sparse young sample they used. Instead, we fit a polynomial to the data
\begin{equation}
\begin{split}\label{eqw:poly}
    {\rm Li\ I\ EqW}=a_0+a_1t+a_2t^2+a_3t^3+a_4t^4+\\
    b_1T+b_2Tt+b_3Tt^2+b_4Tt^3+\\
c_2T^2+c_3T^2t+c_4T^2t^2+\\
d_3T^3+d_4T^3t+
\\e_4T^4
\end{split}
\end{equation}
\noindent where $t$ is $\log_{10}$ age in years, and $T$ is $\log_{10}$\teff\ in K, and the coefficients are listed in Table \ref{tab:coeff}. This relationship is valid for 3200$<$\teff$<$6000 K, and for ages $<$30 Myr.

\begin{deluxetable}{cc|cc}
\tablecaption{Fitted coefficients for Li I estimation
\label{tab:coeff}}
\tabletypesize{\footnotesize}
\tablewidth{\linewidth}
\tablehead{
 \colhead{Coefficient} &
 \colhead{Value} & 
 \colhead{Coefficient} &
 \colhead{Value}
 }
\startdata
$a_0$ & -5148.728 &$b_4$ & 0.4496906 \\
$a_1$ & -523.1078 &$c_2$ & -2422.647\\
$a_2$ & 106.6105 & $c_3$ & 1.814952\\
$a_3$ & -9.375462 &$c_4$ & -0.4258554\\
$a_4$ & 0.2786480 & $d_3$ & 413.0113\\
$b_1$ & 6216.321 &$d_4$ & 0.3173983\\
$b_2$ & 31.22195 &$e_4$ & -26.54814\\
$b_3$ & -5.656809 \\
\enddata
\end{deluxetable}

\begin{figure}
\epsscale{1.1}
\plotone{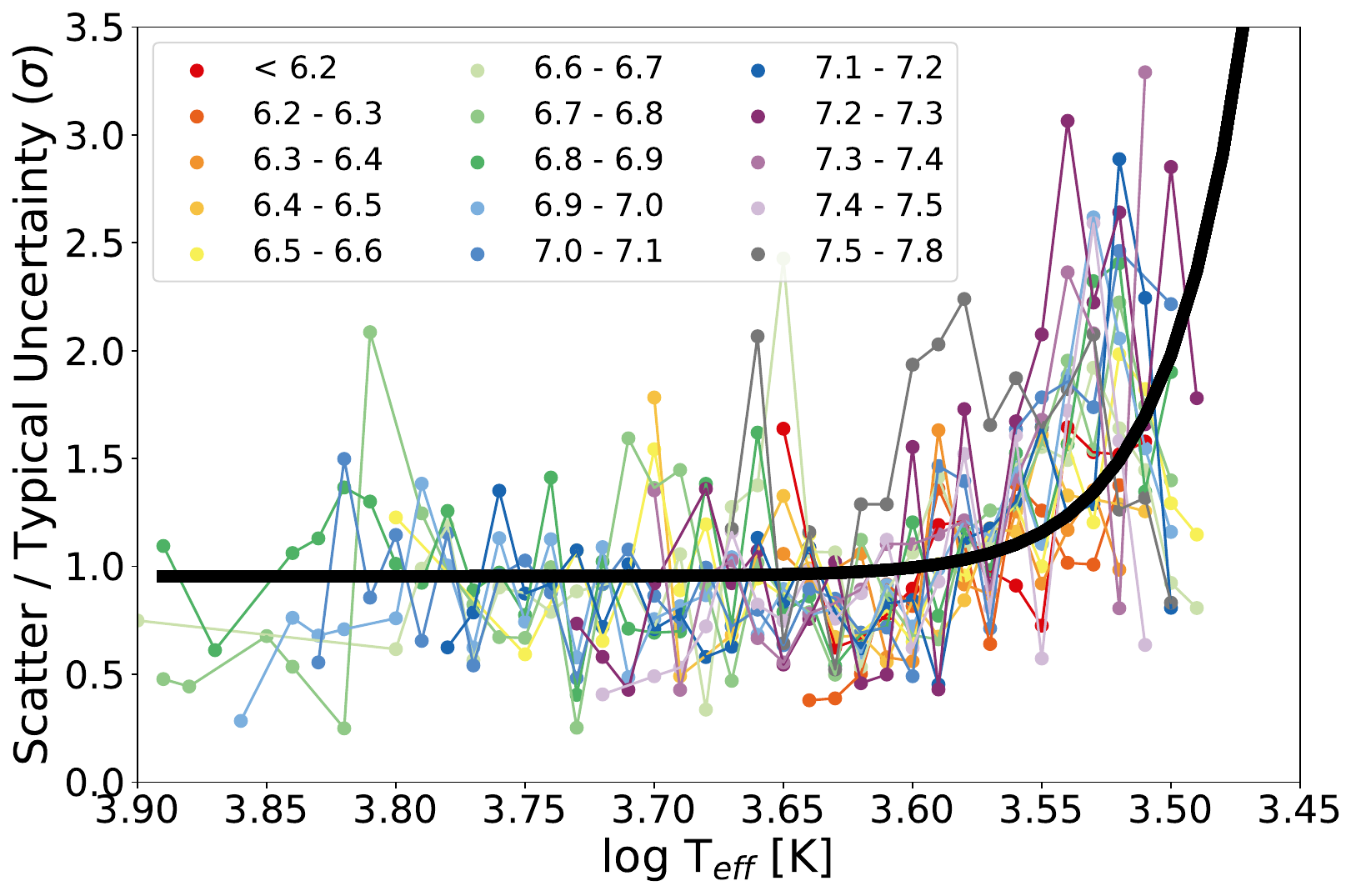}
\caption{Ratio between the typical scatter in Li I in each age and \teff\ bin, and the typical uncertainty in Li I measurements in that bin. The black line shows the fitted average across all age ranges.
\label{fig:scatter}}
\end{figure}

We find that, at all age ranges, at \teff$>$4500 K, the scatter is statistically reproducible by the measurement uncertainties, but at cooler \teff, the scatter becomes almost 2 times larger than the uncertainties (Figure \ref{fig:scatter}). This supports the conclusions of \citet{binks2022} that at low \teff\ Li I abundance in the chromosphere is influenced by the variability in the spot coverage of stars (as opposed to erroneous age measurements that would lead to unrealistic age spread in a given cluster).

\begin{figure}
\epsscale{1.1}
\plotone{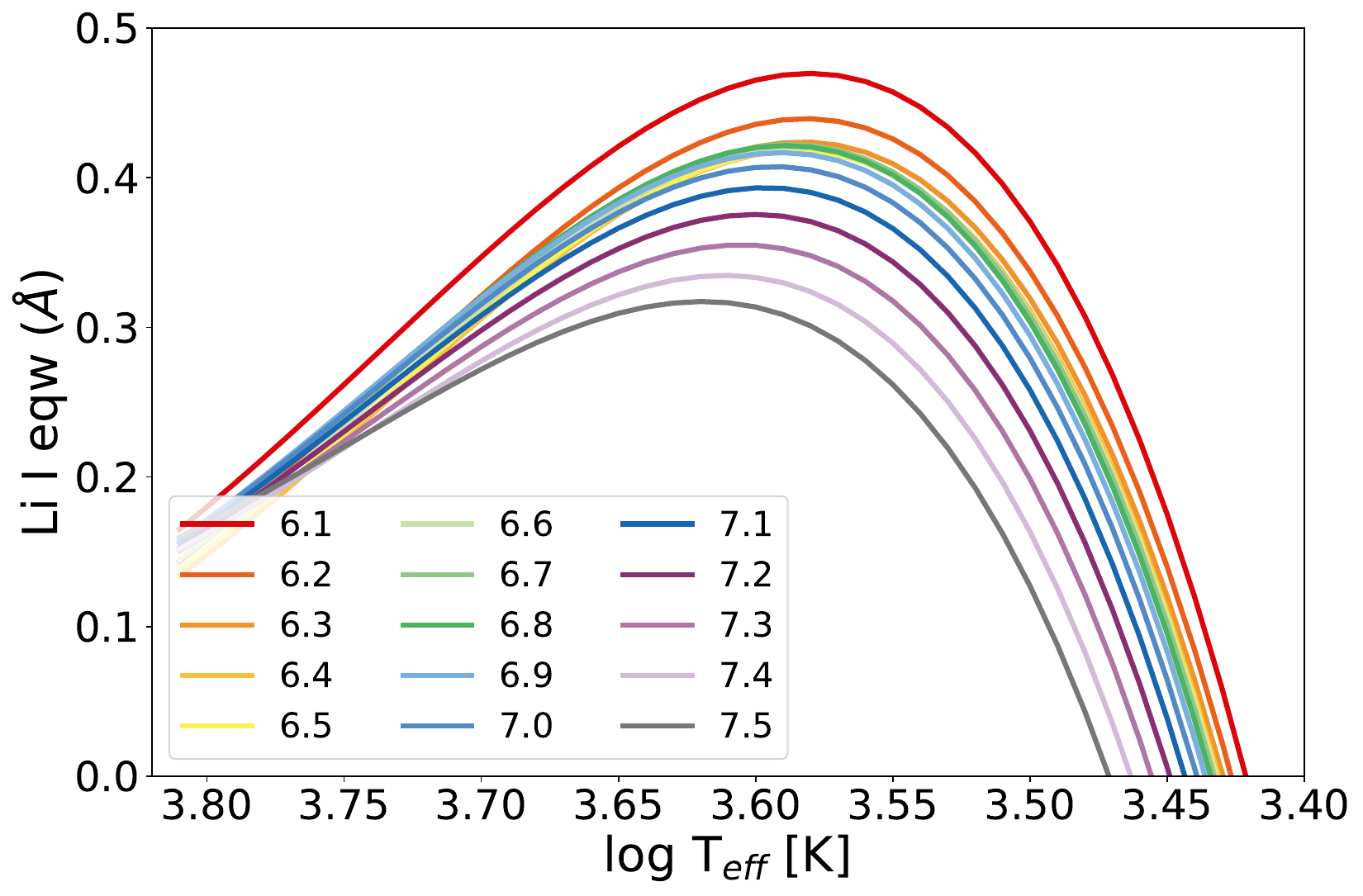}
\caption{Evolution of Li I as a function of age, as characterized by Equation \ref{eqw:poly}.
\label{fig:progression}}
\end{figure}

Comparing Li I at the neighboring age bins, we find that there is a significant depletion from $\sim$1.5 to 3 Myr, after which it stalls until 10 Myr, and beyond that point the abundances once again continue to decline (Figure \ref{fig:progression}. This is apparent both in the fitted expression for estimating average Li I EqW, as well as in comparing the medians computed from the data directly. It is not entirely clear what causes this stalling.

\section{Conclusions}
We present a set of measurements of equivalent widths youth-sensitive lines in optical BOSS \& LAMOST spectra, including lines such as H$\alpha$ and Li I, as well as a number of others. These measurements were taken using a newly developed data-driven pipeline LineForest that was trained on manual measurements of EqWs of $\sim$3500 stars. Although there is a significant emphasis in using these data for characterizing young stars, LineForest is capable of measuring EqWs of these lines in all stellar spectra.

Combining EqWs, AbWs, as well as \teff, \logg, and colors, we developed a classifier that identifies stars younger than a few 10s of Myr. This classifier is most effective in identifying late type YSOs, however, it is capable of confirming youth in the bulk of early-type stars, as well as some of the solar-type stars.

We observe a number of differences in the line properties of YSOs and the more evolved field stars. In particular, activity-sensitive lines (such as the lines of Ca, and the lines of H) tend to show stronger emission (or weaker absorption) in WTTSs than in magnetically active field dwarfs. This is consistent with the decrease in magnetic activity strength as stars become older. We also observe emission lines that are attributable to accretion across a number of elements, some of which are only seen in late-type stars, while others are prominent in stars of all masses.

We examine the decrement of the H lines in the Balmer series for CTTSs, and use these data to estimate the properties of the accretion stream. While the results are preliminary due to the further need to correct the lines for processes such as veiling or extinction, we do find Temp. and log n of the streams to be somewhat consistent with previous studies.

Finally, we examine the evolution of Li I as a function of age. We find that Li I abundance decreases strongly in the first couple of Myr, but then it stalls until the age of 10 Myr, after which point it continues to deplete. We also find that late-type stars appear to exhibit significant scatter in Li I EqWs regardless of the age of the star, and that Li I abundances may be strongly influenced by the variability in the magnetic activity.




\software{TOPCAT \citep{topcat}, BOSS Net \citep{bossnet}, LineForest \citep{lineforest}}

\acknowledgments

CGRZ acknowledges support from projects CONACYT CB2018 A1-S-9754 AND UNAM-PAPIIT AG101723.

Funding for the Sloan Digital Sky Survey V has been provided by the Alfred P. Sloan Foundation, the Heising-Simons Foundation, the National Science Foundation, and the Participating Institutions. SDSS acknowledges support and resources from the Center for High-Performance Computing at the University of Utah. The SDSS web site is \url{www.sdss.org}.

SDSS is managed by the Astrophysical Research Consortium for the Participating Institutions of the SDSS Collaboration, including the Carnegie Institution for Science, Chilean National Time Allocation Committee (CNTAC) ratified researchers, the Gotham Participation Group, Harvard University, Heidelberg University, The Johns Hopkins University, L'Ecole polytechnique f\'{e}d\'{e}rale de Lausanne (EPFL), Leibniz-Institut f\"{u}r Astrophysik Potsdam (AIP), Max-Planck-Institut f\"{u}r Astronomie (MPIA Heidelberg), Max-Planck-Institut f\"{u}r Extraterrestrische Physik (MPE), Nanjing University, National Astronomical Observatories of China (NAOC), New Mexico State University, The Ohio State University, Pennsylvania State University, Smithsonian Astrophysical Observatory, Space Telescope Science Institute (STScI), the Stellar Astrophysics Participation Group, Universidad Nacional Aut\'{o}noma de M\'{e}xico, University of Arizona, University of Colorado Boulder, University of Illinois at Urbana-Champaign, University of Toronto, University of Utah, University of Virginia, Yale University, and Yunnan University.

This work has made use of data from the European Space Agency (ESA)
mission {\it Gaia} (\url{https://www.cosmos.esa.int/gaia}), processed by
the {\it Gaia} Data Processing and Analysis Consortium (DPAC,
\url{https://www.cosmos.esa.int/web/gaia/dpac/consortium}). Funding
for the DPAC has been provided by national institutions, in particular
the institutions participating in the {\it Gaia} Multilateral Agreement.

\bibliographystyle{aasjournal.bst}
\bibliography{main.bbl}

\end{document}